\documentclass[           
               showpacs,            
               preprintnumbers,     
               aps,                 
               prd,                 
               superscriptaddress,      
               nofootinbib,         
               tightenlines,        
               floats,floatfix      
               ]{revtex4}

\usepackage{amsmath}
\usepackage{amssymb}
\usepackage{graphicx}
\newcommand{\Real}{\mathbb{R}}

\begin{document}


\title{Instability of wormholes supported by a ghost scalar field. II. 
Nonlinear evolution}


\author{J. A. Gonz\'alez}
\affiliation{Instituto de F\'{\i}sica y Matem\'{a}ticas, Universidad
              Michoacana de San Nicol\'as de Hidalgo. Edificio C-3, Cd.
              Universitaria, A. P. 2-82, 58040 Morelia, Michoac\'{a}n,
              M\'{e}xico.}

\author{F. S. Guzm\'an}
\affiliation{Instituto de F\'{\i}sica y Matem\'{a}ticas, Universidad
              Michoacana de San Nicol\'as de Hidalgo. Edificio C-3, Cd.
              Universitaria, A. P. 2-82, 58040 Morelia, Michoac\'{a}n,
              M\'{e}xico.}

\author{O. Sarbach}
\affiliation{Instituto de F\'{\i}sica y Matem\'{a}ticas, Universidad
              Michoacana de San Nicol\'as de Hidalgo. Edificio C-3, Cd.
              Universitaria, A. P. 2-82, 58040 Morelia, Michoac\'{a}n,
              M\'{e}xico.}


\date{\today}


\begin{abstract}
We analyze the nonlinear evolution of spherically symmetric wormhole
solutions coupled to a massless ghost scalar field using numerical
methods. In a previous article we have shown that static wormholes
with these properties are unstable with respect to linear
perturbations. Here, we show that depending on the initial
perturbation the wormholes either expand or decay to a Schwarzschild
black hole. We estimate the time scale of the expanding solutions and
the ones collapsing to a black hole and show that they are consistent
in the regime of small perturbations with those predicted from
perturbation theory. In the collapsing case, we also present a
systematic study of the final black hole horizon and discuss the
possibility for a luminous signal to travel from one universe to the
other and back before the black hole forms. In the expanding case, the
wormholes seem to undergo an exponential expansion, at least during
the run time of our simulations.
\end{abstract}




\maketitle



\section{Introduction}

Motivated by spectacular predictions such as interstellar travel and
time machines \cite{mMkT88,mMkTyU88,Visser-Book} there has been a lot
of interest in wormhole physics over the last years. Although wormhole
geometries are speculative because they require exotic matter in order
to represent consistent solutions of Einstein's field equations,
recent cosmological observations (see \cite{ghostcosmo1} for a review)
may suggest the existence of phantom energy which could in principle
be used to support a wormhole \cite{sS05,fL05,fL05b}.

Assuming that stationary wormholes do exist within a particular matter
model, a relevant question is whether or not they are stable with
respect to perturbations. Motivated by this question, in a previous
paper \cite{jGfGoS-inprep1} we performed a stability analysis of
static, spherically symmetric wormhole solutions supported by a
minimally coupled massless ghost scalar field. As we proved, such
wormhole solutions are unstable with respect to linear, spherically
symmetric perturbations of the metric and the scalar field. More
specifically, we showed that each such solution possesses a unique
unstable mode which grows exponentially in time, is everywhere regular
and decays exponentially in the asymptotic region. The time scale
associated to this mode is of the order of the areal radius of the
wormhole throat divided by the speed of light.

The aim of this paper is twofold. First, we verify the wormhole
instability predicted by the linear stability analysis carried out in
our previous paper. For this, we first construct analytically initial
data which describes a nonlinear perturbation of the data induced by
the equilibrium configuration on a static slice. The time evolution of
this data is obtained by numerically integrating the nonlinear field
equations and we show that small initial perturbations do not decay to
zero but grow in time and eventually kick the wormhole throat away
from its equilibrium configuration. The growth of the departure of the
areal radius of the throat from its equilibrium value as a function of
proper time at the throat is computed for different parameters of the
initial perturbation and shown to agree well with the time scale
predicted by the perturbation calculation.

Our second goal is to determine the end state of the nonlinear
evolution. We find that depending on the details of the initial
perturbation the wormhole either starts expanding or collapsing. In
the collapsing case, we observe the formation of an apparent horizon.
By computing geometric quantities at the apparent
horizon, we obtain a strong indication that the apparent horizon
eventually settles down to the event horizon of a Schwarzschild black
hole. By analyzing the convergence of light rays near the apparent
horizon we obtain an estimate for the location of the event horizon
and find that although asymptotically it seems to agree with the
location of the apparent horizon it lies inside it at early times.
Despite the rapid collapse, we find that it is possible for a photon
to travel from one universe to the other and back without falling into
the black hole. In the expanding case, we do not observe any apparent
horizons and our simulations suggest that the wormhole expands
forever. In particular, for wormholes which are reflection symmetric
we find that the areal radius at the point of reflection symmetry
grows exponentially as a function of proper time. Furthermore, we
point out some problems with defining the wormhole throat in an
invariant way. Specifically, we show that the definition of the throat
as a minimum of the areal radius over each time-slice does not yield
an invariant three surface, and we give an example showing that this
three surface depends on the slicing condition.

Furthermore, we perform a careful study for the relation between the
radius of the apparent horizon of the final black hole and the
amplitude of the perturbation. By varying this amplitude, three
regions are found: i) positive values of the amplitude result in the
collapse to a black hole, ii) negative values of the amplitude lying
sufficiently close to zero trigger the explosion of the solution, iii)
a second threshold for the amplitude is found on the negative real
axis below which the wormholes collapse to a black hole.

This paper is organized as follows. In section~\ref{sec:cauchy} we
formulate the Cauchy problem for obtaining the nonlinear time
evolution of spherically symmetric configurations, review the static
wormhole solutions and describe our method for obtaining initial
perturbations thereof. Our numerical implementation and the
diagnostics we use in order to test our code and analyze the data
obtained from the evolution are described in
section~\ref{sec:diagnostics}. In section~\ref{sec:collapse} we
analyze the collapsing case and indicate that the final state is a
Schwarzschild black hole. By analyzing the resulting spacetime 
diagram we also mention different scenarios for a light ray to travel 
from one universe to the other and back. The expanding case is analyzed 
in section~\ref{sec:expanding} and the study for the relation between 
the final state and the amplitude of the perturbation is carried out in
section~\ref{sec:finalstate}. Finally, we summarize our findings and
draw conclusions in section~\ref{sec:conclusions}.

Some of the conclusions presented in this paper have been presented in
Ref. \cite{hSsH02} for the wormholes with zero ADM masses based on a
numerical method using null coordinates.


\section{The Cauchy problem for time-dependent solutions}
\label{sec:cauchy}

In this section we formulate Einstein's equations coupled to a
massless ghost scalar field in a suitable form for the numerical
integration of spherically symmetric wormhole solutions. We use 
geometrized units for which the speed of light and Newton's constant are 
equal to one. We write the metric in the form
\begin{equation}
ds^2 = -e^{2d} dt^2 + e^{2a} dx^2 
 + e^{2c}\left( d\vartheta^2 + \sin^2\vartheta\; d\varphi^2 \right),
\label{Eq:SphericalMetric}
\end{equation}
where the functions $d = d(t,x)$, $a = a(t,x)$ and $c = c(t,x)$ depend
only on the time coordinate $t$ and the radial coordinate
$x$. Similarly, we assume that the scalar field $\Phi = \Phi(t,x)$
only depends on $t$ and $x$. The coordinate $x$ ranges over the whole
real line, where the two regions $x\to\infty$ and $x\to -\infty$
describe the two asymptotically flat ends. In particular, we require
that the two-manifold $(\tilde{M},\tilde{g}) = (\Real^2,-e^{2d} dt^2 +
e^{2a} dx^2)$ is regular and asymptotically flat at $x\to \pm\infty$,
and that the areal radius $r = e^c$ is strictly positive and
proportional to $|x|$ for large $|x|$. For this reason, it is
convenient to replace $c$ by $\bar{c}$ where
\begin{equation}
c = \bar{c} + \frac{1}{2}\log(x^2 + b^2)
\end{equation}
and $b > 0$ is a parameter. 
Next, we find it convenient to choose the following family of gauge
conditions:
\begin{equation}
d = a + 2\lambda\bar{c},
\label{Eq:SlicingCondition}
\end{equation}
which relates the lapse $e^d$ to the functions $a$ and $\bar{c}$
parametrizing the three-metric. Here, $\lambda$ is a parameter that
may take any real value in principle, but for our simulations we will
consider the choices $\lambda=0$ and $\lambda=1$. The metric now takes
the form
\begin{equation}
ds^2 = e^{2a}\left( -e^{4\lambda\bar{c}} dt^2 + dx^2 \right) 
 + (x^2 + b^2) 
   e^{2\bar{c}}\left( d\vartheta^2 + \sin^2\vartheta\; d\varphi^2 \right).
\end{equation}

\subsection{Evolution and constraint equations}

Einstein's equations yield the Hamiltonian constraint ${\cal H} :=
-e^{a-d} G_{tt} + \kappa[e^{a-d}\Phi_t^2 + e^{d-a}\Phi_x^2]/2 = 0$,
the momentum constraint ${\cal M} := -(R_{xt} - \kappa\Phi_t\Phi_x) =
0$ and the evolution equations $R_{\vartheta\vartheta} =
R_{\varphi\varphi} = 0$ and $e^{d-a}( R_{xx} - \kappa\Phi_x^2 ) + (1 +
\lambda){\cal H} = 0$, where here $G_{\mu\nu}$ and $R_{\mu\nu}$ refer,
respectively, to the components of the Einstein and Ricci tensor,
$\kappa=-8\pi $ and $\Phi_t :=
\partial_t\Phi$ and $\Phi_x := \partial_x\Phi$. These evolution
equations, together with the wave equation for $\Phi$ yield the
following coupled system of nonlinear wave equations for the
quantities $a$, $\bar{c}$ and $\Phi$:
\begin{eqnarray}
&& \partial_t\left( e^{-2\lambda\bar{c}} a_t \right) 
 - \partial_x\left[ e^{2\lambda\bar{c}} 
 \left( a_x - \frac{2\lambda x}{x^2+b^2} \right) \right]
 - e^{-2\lambda\bar{c}}\left[ 
 (1+\lambda)\bar{c}_t^2 + 2\lambda a_t\bar{c}_t \right]
 + e^{2\lambda\bar{c}}
   \left[ (1 + 3\lambda)c_x^2 - 2\lambda c_x(a_x + 2\lambda\bar{c}_x) \right]
\nonumber\\
 &&\qquad -(1+\lambda)\frac{e^{2a + 2(\lambda-1)\bar{c}}}{x^2+b^2}
 + \frac{\kappa}{2}\left[
  (\lambda+1) e^{-2\lambda\bar{c}} \Phi_t^2 
+ (\lambda-1) e^{2\lambda\bar{c}} \Phi_x^2 \right] = 0,
\label{Eq:Evol1}\\
&& \partial_t\left( e^{2(1-\lambda)\bar{c}}\bar{c}_t \right) 
 - \frac{1}{x^2+b^2}\,
   \partial_x\left[ e^{2(1+\lambda)\bar{c}}
   \left( (x^2+b^2)\bar{c}_x + x \right) \right]
 + \frac{e^{2(a+\lambda\bar{c})}}{x^2+b^2}
 = 0,
\label{Eq:Evol2}\\
&& \partial_t\left( e^{2(1-\lambda)\bar{c}}\Phi_t \right) 
 - \frac{1}{x^2 + b^2}\,
   \partial_x\left[ (x^2+b^2) e^{2(1+\lambda)\bar{c}}\Phi_x \right] = 0,
\label{Eq:Evol3}
\end{eqnarray}
which is subject to the constraints
\begin{eqnarray}
{\cal H} &=& e^{2\lambda\bar{c}}( 2c_{xx} + 3c_x^2 - 2a_x c_x ) 
 - e^{-2\lambda\bar{c}} c_t(c_t + 2a_t) 
 - \frac{e^{2a + 2(\lambda-1)\bar{c}}}{x^2 + b^2} 
 + \frac{\kappa}{2} \left[ e^{-2\lambda\bar{c}} \Phi_t^2 
 + e^{2\lambda\bar{c}} \Phi_x^2 \right] = 0, \label{Eq:Ham}\\
{\cal M} &=& 2c_{tx} + 2c_t\left( c_x - 2\lambda\bar{c}_x - a_x \right) 
 - 2a_t c_x + \kappa\,\Phi_t\Phi_x = 0. \label{Eq:Mom}
\end{eqnarray}
For $\lambda=0$ we see that the principal part of the evolution
equations is given by the flat wave operator $\partial_t^2 -
\partial_x^2$. Therefore, the characteristic lines coincide with the
radial null rays which for $\lambda=0$ are given by the straight lines
$t\pm x = const.$ Therefore, a property of the gauge condition
(\ref{Eq:SlicingCondition}) with $\lambda=0$ is that it cannot lead to
shock formations due to the crossing of characteristics. We found this
gauge to be convenient for studying expanding wormholes. For the
collapsing case, on the other hand, we found the gauge
(\ref{Eq:SlicingCondition}) with $\lambda = 1$ more useful since it
seems to be avoiding the singularity after the black hole forms.

\subsection{Propagation of the constraints}

It can be shown that the Bianchi identities and the evolution
equations~(\ref{Eq:Evol1},\ref{Eq:Evol2},\ref{Eq:Evol3}) imply that
the constraint variables ${\cal H}$ and ${\cal M}$ satisfy a linear
evolution system of the form
\begin{eqnarray}
\partial_t{\cal H} &=& e^{2\lambda\bar{c}}\partial_x{\cal M} + l.o.,\\
\partial_t{\cal M} &=& e^{2\lambda\bar{c}}\partial_x{\cal H} + l.o.,
\end{eqnarray}
where $l.o.$ are lower order terms which depend only on ${\cal H}$ and
${\cal M}$ but not their derivatives. For a solution on the unbounded
domain $-\infty < x < \infty$ with initial data satisfying ${\cal H} =
{\cal M} = 0$ it follows that the constraints are automatically
satisfied everywhere and for each $t > 0$. Therefore, it is sufficient
to solve the constraints initially. If artificial time-like boundaries
are imposed, this statement is still true provided suitable boundary
conditions are specified. For example, imposing the momentum
constraint ${\cal M} = 0$ at the boundary ensures that the constraints
remain satisfied if so initially.

\subsection{Boundary conditions}

Since we have three wave equations, at the artificial boundaries we 
apply outgoing wave boundary conditions for the three fields 
$a$, $\bar{c}$ and $\Phi$. We assume that these fields behave like 
spherical waves far away from the origin, {\it i.e.} if $x>0$:
\begin{equation}
f(t,x) = \frac{g(x-vt)}{x}\; ,
\end{equation}
where $v=e^{2\lambda \bar{c}}$ is the speed of propagation.  
In order to apply the boundary conditions, we use the following
equation,
\begin{equation}
\frac{1}{v}\partial_t f + \partial_x f + \frac{f}{x} = 0.
\label{Eq:OutgoingBC}
\end{equation}
We use a similar procedure when $x<0$. It is clear that such
conditions do not preserve the constraints and consequently, the
solution will be contaminated with a constraint-violating pulse
traveling inwards in the numerical domain. In order to avoid the
contamination of the analyzed data, we push the boundaries far from
the throat such that the region where we extract physics is causally
disconnected from the boundaries.

\subsection{Static wormhole solutions}

In the static case, all wormhole solutions can be found analytically
\cite{hE73,kB73,cA02} and are given by the following expressions
\begin{eqnarray}
d &=& -a = \gamma_1\arctan\left( \frac{x}{b} \right),
\label{Eq:StaticWormHole1}\\
\bar{c} &=& -\gamma_1\arctan\left( \frac{x}{b} \right),
\label{Eq:StaticWormHole2}\\
\Phi &=& \Phi_1\arctan\left( \frac{x}{b} \right),
\label{Eq:StaticWormHole3}
\end{eqnarray}
where the parameters $\Phi_1$ and $\gamma_1$ are subject to the
condition $-\kappa\Phi_1^2 = 2(1 + \gamma_1^2)$. The wormhole throat
is located at $x_{throat} = \gamma_1 b$ and has areal radius
$r_{throat} = b\sqrt{1 + \gamma_1^2}
e^{-\gamma_1\arctan(\gamma_1)}$. The ADM masses at the two
asymptotically flat ends $x\to\pm\infty$ are $m_\infty =
b\gamma_1\exp(-\gamma_1\pi/2)$ and $m_{-\infty} =
-b\gamma_1\exp(\gamma_1\pi/2)$, respectively (see
Ref. \cite{jGfGoS-inprep1} for a derivation and details). In
particular, $\gamma_1 = 0$ yields wormhole solutions with zero ADM
mass at both ends while in all other cases the ADM masses at the two
ends have opposite signs.

\subsection{Initial data describing a perturbed wormhole}

In order to study the nonlinear stability of the static solutions
described by
equations~(\ref{Eq:StaticWormHole1},\ref{Eq:StaticWormHole2},\ref{Eq:StaticWormHole3})
we construct initial data corresponding to an initial perturbation of
a static solution. For simplicity, we assume that the initial slice is
time-symmetric in which case the momentum constraint ${\cal M} = 0$ is
automatically satisfied. Therefore, we only need to solve the
Hamiltonian constraint which, for $a_t = c_t = \Phi_t = 0$ simplifies
to
\begin{equation}
2\bar{c}_{xx} + 3\bar{c}_x^2 - 2a_x\bar{c}_x 
 + \frac{2x}{x^2 + b^2}(3\bar{c}_x - a_x)
 + \frac{1 - e^{2(a - \bar{c})}}{x^2 + b^2} 
 + \frac{b^2}{(x^2 + b^2)^2}
 = -\frac{\kappa}{2} \Phi_x^2.
\label{Eq:HamTimeSym}
\end{equation}
Since the unperturbed wormhole solution is known in analytic form a
simple way of obtaining initial data representing a perturbation
thereof is to perturb the metric quantities $a$ and $\bar{c}$ by hand
in such a way that the left-hand side of
equation~(\ref{Eq:HamTimeSym}) is nonnegative and to solve
equation~(\ref{Eq:HamTimeSym}) for $\Phi_x$. Since the left-hand side
of Equation~(\ref{Eq:HamTimeSym}) is everywhere positive for the
analytic solution, it remains positive at least for small enough
perturbations. Moreover, it is clear that any time-symmetric and
spherically symmetric initial data can be obtained by this method.
Next, we notice that it is sufficient to consider the case where the
metric component $a$ is unperturbed. Indeed, a perturbation of $a$ may
be absorbed by a redefinition of the coordinate $x$ which, at the
physical level, does not change the initial data and its Cauchy
development. Therefore, it is sufficient to perturb the quantity
$\bar{c}$ which is related to the areal radius. For this work, we
choose to perturb it with a Gaussian pulse. More precisely, we choose
\begin{eqnarray}
a_{\rm pert} &=& a_{\rm static}\; ,
\label{Eq:a_perturbed}\\
\bar{c}_{\rm pert} &=& \bar{c}_{\rm static} 
 + \varepsilon_c e^{-(x-x_c)^2/\sigma_c^2}\, ,
\label{Eq:c_perturbed}
\end{eqnarray}
where $\varepsilon_c$ is the amplitude, $x_c$ the center and $\sigma_c$
is related to the width $w$ at half maximum of the pulse through $w =
2\sqrt{\log 2}\,\sigma_c$. As will be shown later, different values of
$\varepsilon_c$ and $\sigma_c$ produce two different scenarios: the
collapse of the wormhole to a black hole or a rapid expansion.  Notice
that because of the exponential decay of the perturbation as
$|x|\to\pm\infty$, the ADM masses of the perturbed solution is equal
to the ADM masses of the unperturbed, static solution.

\section{Implementation and diagnostics}
\label{sec:diagnostics}

In this section we describe our numerical method and different tools
used to analyze the data obtained from the simulations. The numerical
method is based on a second order centered finite differences
approximation of the evolution
equations~(\ref{Eq:Evol1}-\ref{Eq:Evol3}) and the constraint
equations~(\ref{Eq:Ham},\ref{Eq:Mom}). We only perform the evolution
on a finite domain with artificial boundaries at a finite value of $x$
where we implement the radiative-type boundary condition
(\ref{Eq:OutgoingBC}) with an accuracy of second order for the
evolution variables $a$, $\bar{c}$ and $\Phi$. A method of lines using
the third order Runge-Kutta integrator is adopted. Throughout the
evolution, we monitor the constraint variables ${\cal H}$ and ${\cal
M}$ and check that they converge to zero as resolution is increased.

From now on we rescale the coordinate $x$ in such a way that $b=1$.

\subsection{Unperturbed wormhole}

\begin{figure}[ht]
\includegraphics[width=8cm]{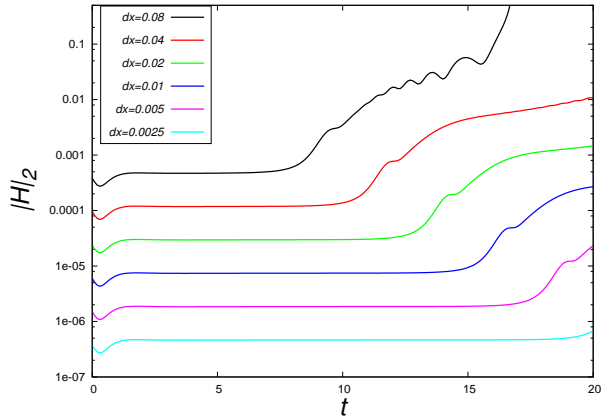}
\caption{The $L_2$ norm of the Hamiltonian constraint variable ${\cal
H}$ as a function of coordinate time for an unperturbed wormhole with
gauge parameter $\lambda=1$. Different resolutions are shown. As is
apparent from the plot at each fixed time, the error decreases with
increasing resolution.}
\label{fig:ham_no_pert_rms}
\end{figure}

As a test, using the implementation aforementioned we perform a series
of simulations for the unperturbed, static wormholes. We start a
simulation with the massless case $\gamma_1=0$, for which it is
expected that the fields $a,\bar{c},\Phi$ remain
time-independent. However, discretization errors are enough to trigger
a non-trivial time dependence of these functions. We have verified
that when increasing the resolution, the numerical error remains small
for a larger period of time and the numerical solution converges to
the exact, static solution with second order. In
figure~\ref{fig:ham_no_pert_rms} we show the convergence of the
constraint variable ${\cal H}$ for a wormhole using the gauge
parameter $\lambda=1$.

\subsection{Extremal and marginally trapped surfaces}

Let us consider a $t = const$ hypersurface $\Sigma_t$. There are two
different types of interesting two-surfaces in this hypersurface: The
first are {\em extremal surfaces} which consist of two-surfaces with a
local extrema of their area. If $s^i$ denotes the unit normal to this
surface, this means that the divergence $D_i s^i$ vanishes, where $D$
refers to the covariant derivative associated to the induced
three-metric on $\Sigma_t$. The second type of two-surfaces of
interest are {\em marginally trapped surfaces}. Denoting by $l^\alpha$
and $k^\alpha$ the future-pointing outward and inward null-vectors
orthogonal to this two-surface, a marginally trapped surface is
defined by the requirement that the expansion $\theta_-$ along
$k^\alpha$ is strictly negative while the expansion $\theta_+$ along
$l^\alpha$ vanishes. In terms of the induced metric
$\gamma_{\alpha\beta}$ on the two-surface, this means that
\begin{displaymath}
\theta_- := \gamma_{\alpha\beta}\nabla^\alpha k^\beta < 0, \qquad
\theta_+ := \gamma_{\alpha\beta}\nabla^\alpha l^\beta = 0.
\end{displaymath}
Notice that a rescaling of $k^\alpha$ or $l^\alpha$ by a positive
function does not affect this definition. On the other hand, for
wormhole topologies the notions ``outward'' and ``inward'' are
observer-dependent: they depend on which asymptotic end these
vectors are viewed from.

With respect to $t=const$ slices of the spherically symmetric metric
(\ref{Eq:SphericalMetric}) we have $s^i\partial_i = e^{-a}\partial_x$,
$k^\alpha\partial_\alpha = e^{-d}\partial_t - e^{-a}\partial_x$,
$l^\alpha\partial_\alpha = e^{-d}\partial_t + e^{-a}\partial_x$ (with
respect to the asymptotic end $x\to\infty$), and
\begin{equation}
D_i s^i = 2 e^{-a} c_x\; ,\qquad
\theta_- = 2(e^{-d}c_t - e^{-a} c_x), \qquad
\theta_+ = 2(e^{-d}c_t + e^{-a} c_x).
\end{equation}
Therefore, an extremal surface is a sphere $S^2$ for which $c_x = 0$,
and a marginally trapped surface a sphere for which $e^{-d} c_t =
-e^{-a} c_x$ and $c_x > 0$. In particular, according to our
definition, an extremal surface cannot be a marginally trapped
surface. Finally, we define a throat to be an extremal surface which
(out of the many extremal surfaces that might exist) is one with least
area. In a similar way, an apparent horizon is an outermost marginally
trapped surface.  Notice that these concepts are not defined in a
geometrically invariant way. For example, it is known \cite{rWvI91} that
the Schwarzschild spacetime possesses Cauchy surfaces which do not
contain any trapped surfaces and nevertheless come arbitrarily close
to the singularity. Similarly, we will provide a numerical example
below that shows that the three-surface obtained by piling up the
throats in each time slice is not invariant but depends on the time
foliation.

\subsection{Geometric quantities}

Throughout evolution, we monitor the values of the following geometric
quantities:
\begin{eqnarray}
r &=& e^c = \sqrt{x^2 + b^2}\, e^{\bar{c}},
\\
M &=& \frac{r}{2}(1-\tilde{g}^{ab}\tilde{\nabla}_a r\cdot\tilde{\nabla}_b r) 
 = \frac{e^c}{2}\left[ 1 - e^{2c}(-e^{-2d}c_t^2 + e^{-2a}c_x^2) \right],
\label{Eq:MInvariant}\\
L &=& \tilde{g}^{ab}\tilde{\nabla}_a\Phi \cdot \tilde{\nabla}_b\Phi =
-e^{-2d}\Phi_t^2 + e^{-2a}\Phi_x^2,
\label{Eq:LInvariant}
\end{eqnarray}
where quantities with a tilde refer to the two-metric $\tilde{g} =
-e^{2d} dt^2 + e^{2a} dx^2$. As mentioned above, $r = r(t,x)$ is the
areal radius of the sphere at constant $t$ and $x$. $M$ is the
Misner-Sharp mass function \cite{cMdS64} and $L$ is the norm of the
gradient of the scalar field. In particular, the knowledge of these
functions allows the computation of the following curvature scalars:
The Ricci scalar $\tilde{R}$ associated to the two-metric $\tilde{g}$,
the Kretschmann scalar associated to the full metric
$I:=R^{\alpha\beta\gamma\delta} R_{\alpha\beta\gamma\delta}$, and the
square of the Ricci tensor, $J:=R^{\mu\nu} R_{\mu\nu}$. Using
Einstein's field equations one obtains the following expressions,
\begin{eqnarray}
\tilde{R} &=& \frac{4M}{r^3} + \kappa L,\\
I &=& \frac{48M^2}{r^6} + \frac{8\kappa M}{r^3} L + 2\kappa^2 L^2,\\
J &=& \kappa^2 L^2.
\end{eqnarray}

\subsection{The construction of conformal coordinates}
\label{subsec:ConfCoords}

The gauge choice (\ref{Eq:SlicingCondition}) with $\lambda = 0$ has
the property of yielding conformally flat coordinates $(t,x)$ for the
two-metric $\tilde{g} = -e^{2d} dt^2 + e^{2a} dx^2$. In these
coordinates, the null rays are simply given by the straight lines $t
\pm x = const.$

On the other hand, in many simulations, the choice $\lambda = 1$ seems
to work better than the choice $\lambda=0$. In order to compare the
coordinates $(T,X)$, say, constructed with the gauge choice
$\lambda=1$ to the conformally flat coordinates $(t,x)$ obtained with
$\lambda=0$ one can proceed as follows. Suppose we are given to us a
two-metric $\tilde{g} = \tilde{g}_{ab} dx^a dx^b$ with signature
$(-1,1)$. We are interested in finding local coordinates $(t,x)$ such
that
\begin{equation}
\tilde{g} = e^{2a}(-dt^2 + dx^2),
\label{Eq:ConformalCoords}
\end{equation}
with a conformal factor $e^a$. We first note that if
$\tilde{\varepsilon}_{ab}$ denotes the volume element associated to
$\tilde{g}$, the coordinates $t$ and $x$ must satisfy
\begin{equation}
\tilde{\nabla}_a t = \pm\tilde{\varepsilon}_a{}^b\tilde{\nabla}_b x\; ,
\label{Eq:ConfConstraint}
\end{equation}
since by equation~(\ref{Eq:ConformalCoords}) $\tilde{\nabla}_a t$ and
$\tilde{\nabla}_b x$ are orthogonal to each other and their norms are
equal in magnitude. It follows from equation~(\ref{Eq:ConfConstraint})
that $t$ and $x$ must also satisfy the wave equation
\begin{equation}
\tilde{\nabla}^a\tilde{\nabla}_a t = \tilde{\nabla}^a\tilde{\nabla}_a x = 0.
\label{Eq:ConfWaveEq}
\end{equation}

Therefore, in order to construct the coordinates $(t,x)$, we have to
solve equations~(\ref{Eq:ConfWaveEq}) subject to the constraint 
(\ref{Eq:ConfConstraint}). The resulting functions $(t,x)$ indeed
satisfy (\ref{Eq:ConformalCoords}) since
\begin{eqnarray*}
&& -\tilde{g}^{tt} = -\tilde{g}^{ab}(\tilde{\nabla}_a t)(\tilde{\nabla}_b t)
 = \tilde{g}^{ab}(\tilde{\nabla}_a x)(\tilde{\nabla}_b x) = \tilde{g}^{xx},
\\
&& \tilde{g}^{tx} = \tilde{g}^{xt} = 
\tilde{g}^{ab}(\tilde{\nabla}_a t)(\tilde{\nabla}_b x) = 0.
\end{eqnarray*}
The conformal factor is then obtained from $e^a =
1/\sqrt{\tilde{g}^{xx}}$. Notice that equation~(\ref{Eq:ConfWaveEq})
implies that both $C_a := \tilde{\nabla}_a t \pm
\tilde{\varepsilon}_a{}^b\tilde{\nabla}_b x$ and $D_a :=
\tilde{\varepsilon}_a{}^b C_b$ are conserved, i.e. $\tilde{\nabla}^a
C_a = \tilde{\nabla}^a D_a = 0$. Therefore, it is sufficient to solve
the constraint $C_a = 0$ on a Cauchy slice.

For example, suppose that $\tilde{g} = e^{2A}\left( -e^{4\bar{c}} dT^2
+ dX^2 \right)$ as is the case for our simulations with $\lambda =
1$. Then, equations~(\ref{Eq:ConfWaveEq}) yield
\begin{eqnarray*}
\partial_T \left( e^{-2\bar{c}}\partial_T t \right) &=& 
\partial_X \left( e^{2\bar{c}}\partial_X t \right),\\
\partial_T \left( e^{-2\bar{c}}\partial_T x \right) &=& 
\partial_X \left( e^{2\bar{c}}\partial_X x \right),
\end{eqnarray*}
and the initial data for the functions $t = t(T,X)$ and $x = x(T,X)$
satisfies
\begin{eqnarray*}
t(0,X) &=& 0,\\
x(0,X) &=& X,\\
\partial_T t(0,X) &=& e^{2\bar{c}}\partial_X x(0,X) = e^{2\bar{c}(0,X)},\\
\partial_T x(0,X) &=& e^{2\bar{c}}\partial_X t(0,X) = 0.  
\end{eqnarray*}
The last two equations follow from equation~(\ref{Eq:ConfConstraint})
where we have chosen the $+$ sign such that $t$ and $T$ point in the
same direction. The so obtained functions $t = t(T,X)$ and $x =
x(T,X)$ can be used, for instance, to plot a $T=const$ surface in the
$t-x$ diagram. They may also be useful to find the null rays which are
given by the lines with constant $t + x$ or $t - x$.


\section{Collapse to a black hole}
\label{sec:collapse}

\begin{figure}[ht]
\includegraphics[width=8cm]{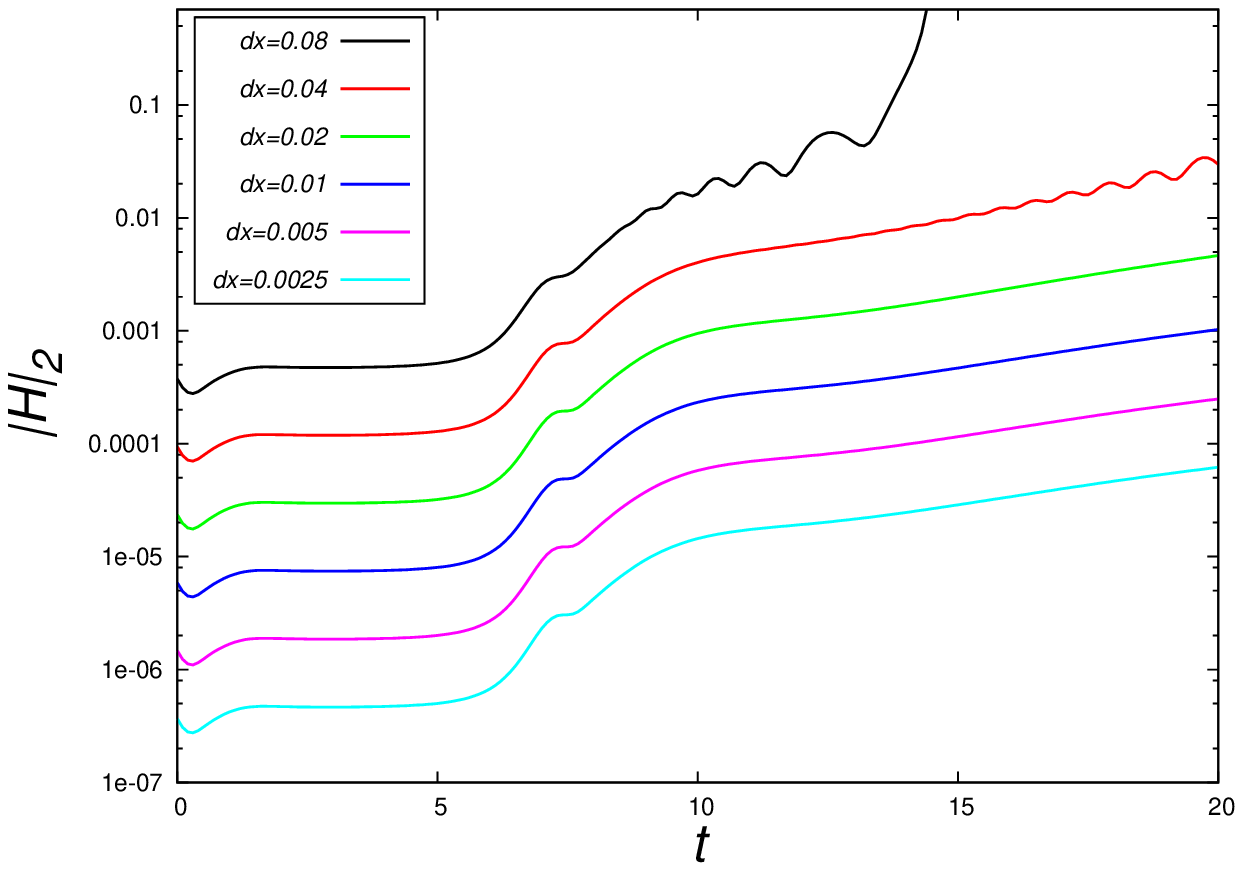}
\includegraphics[width=8cm]{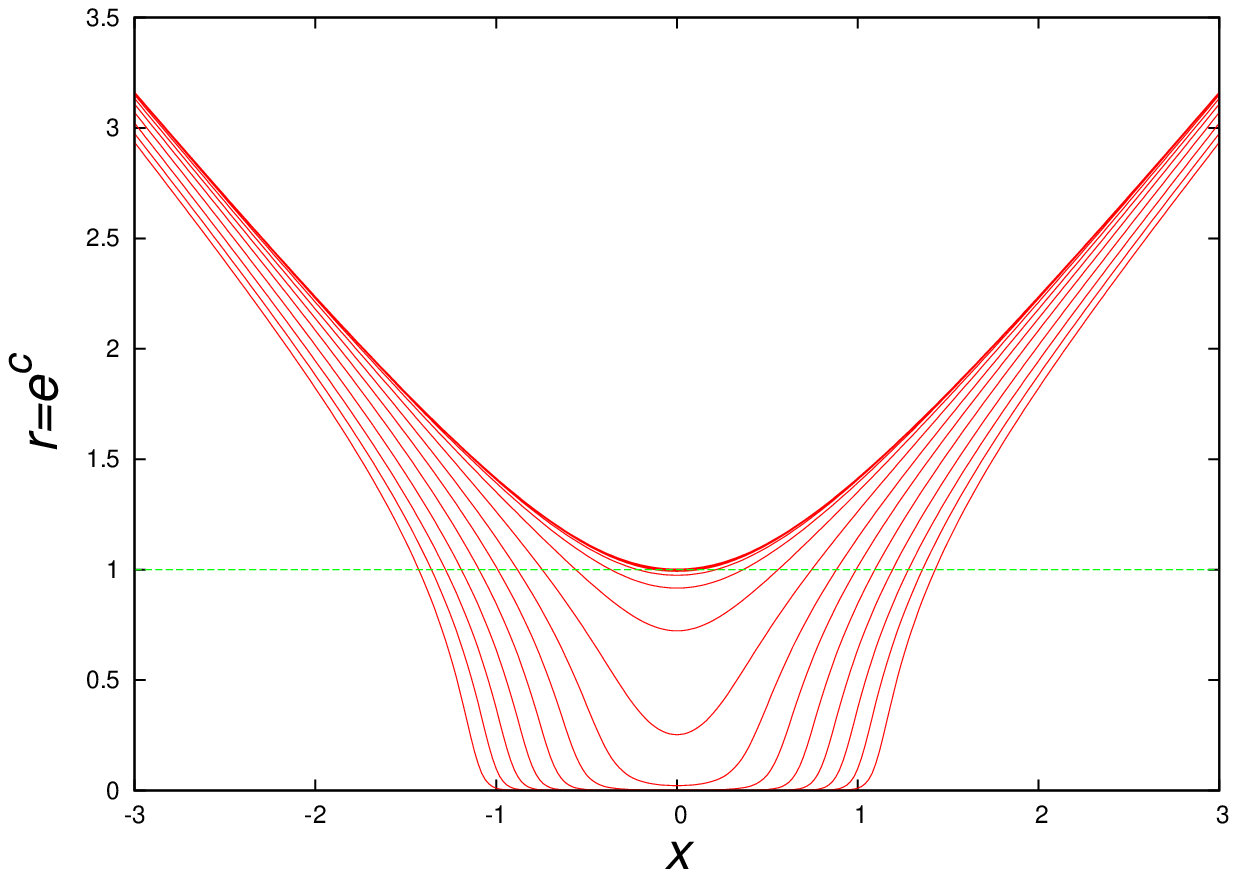}
\caption{(Left panel) $L_2$ norm of the Hamiltonian constraint
variable $H$ as a function of coordinate time for a perturbed wormhole
with parameters $\varepsilon_c = 0.001$ and $\sigma_c = 0.5$ and gauge
parameter $\lambda=1$. Various resolutions are shown. We have checked
that these constraint variables converge to zero with second
order. (Right panel) Areal radius as function of the coordinate $x$ at
different times. In this scenario the value of the throat's areal
radius starts at $r_{throat}=1$ and collapses to zero in finite
coordinate and proper time (see also
figure~\ref{Fig:life-time_Symmetric} below).}
\label{fig:areal_r_pert}
\end{figure}

For simplicity, we start with the evolution of a perturbed, massless
wormhole with the perturbation consisting of a Gaussian pulse as in
(\ref{Eq:c_perturbed}) with $x_c = 0$ so that the perturbation is
centered at the throat. We start with positive values of the amplitude
$\varepsilon_c$ and find that such perturbations induce a collapse of
the throat. The typical behaviour of such a collapsing wormhole is
presented in figures~\ref{fig:areal_r_pert} to
\ref{fig:massive_collpase_r}.

In figure~\ref{fig:areal_r_pert} we show the convergence of the
Hamiltonian constraint as a function of coordinate time and the
evolution of the areal radius which exhibits the collapse of the
wormhole throat.

\subsection{Time scale of the collapse}

In order to check the consistency of our numerical calculations with
the time scale associated to the unstable mode found in our linear
stability analysis, we perform a systematic study of the speed of the
collapse of the throat. For simplicity, we only consider massless
wormholes with a centered perturbation. Our procedure for quantifying
the speed of the collapse is the following:

\begin{itemize}
\item[i)] Compute at each time the throat's location by finding the 
	minimum  of the areal radius. For the reflection symmetric case, 
	this minimum is located at $x=0$.
\item[ii)] Compute the proper time $\tau$ integrating the lapse
	function $e^d$ at the throat.
\item[iii)] Plot the throat's areal radius as a function of proper
time.
\item[iv)] Fit the throat's areal radius as a function of proper time
	to the function $r(\tau) = 1 - e^{-(\tau-p_1)/p_2}$, from the
	initial radius of the throat until a chosen minimum value of
	$r$, called $r_{cut}$. We interpret the parameter $p_2$ as the
	time scale of the solution. The reason to cut off the values
	of $r$ is that we want to compare $p_2$ with the results from
	perturbation theory which are expected to be valid as long as
	the departure from the equilibrium configuration is small.
\item[v)] Repeat the above analysis for fixed initial parameters $x_c
= 0$ and $\sigma_c = 0.5$ and several values of the amplitude
$\varepsilon_c$.
\end{itemize}

\begin{figure*}[htp]
\includegraphics[width=8cm]{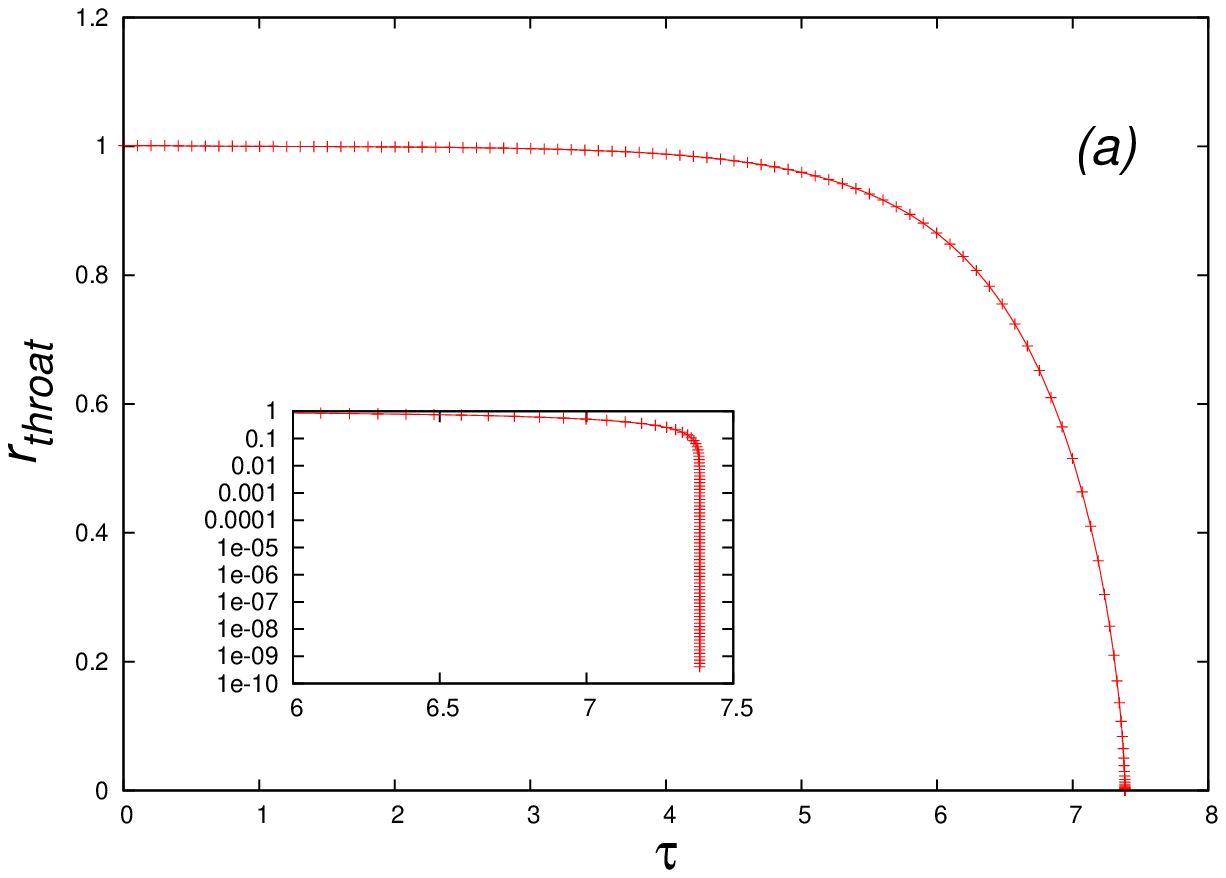}
\includegraphics[width=8cm]{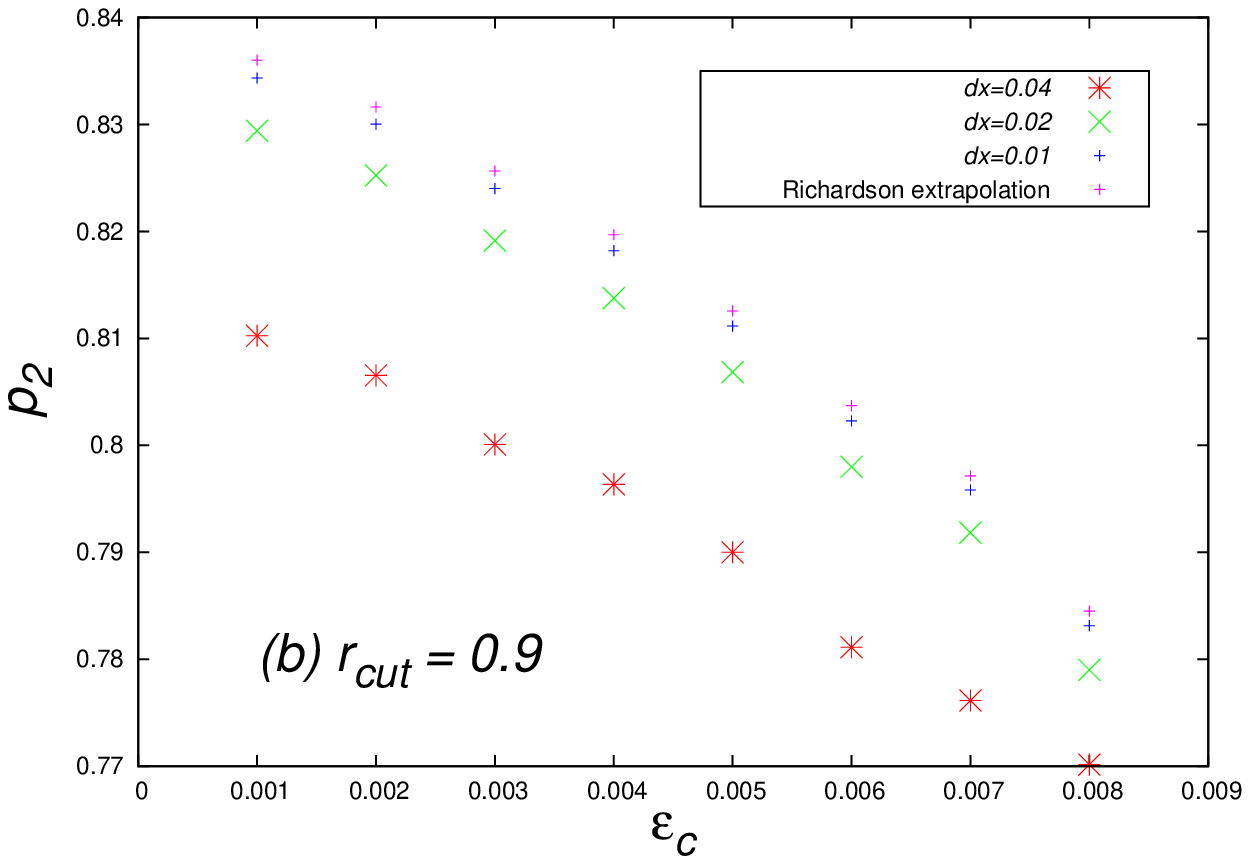}
\includegraphics[width=8cm]{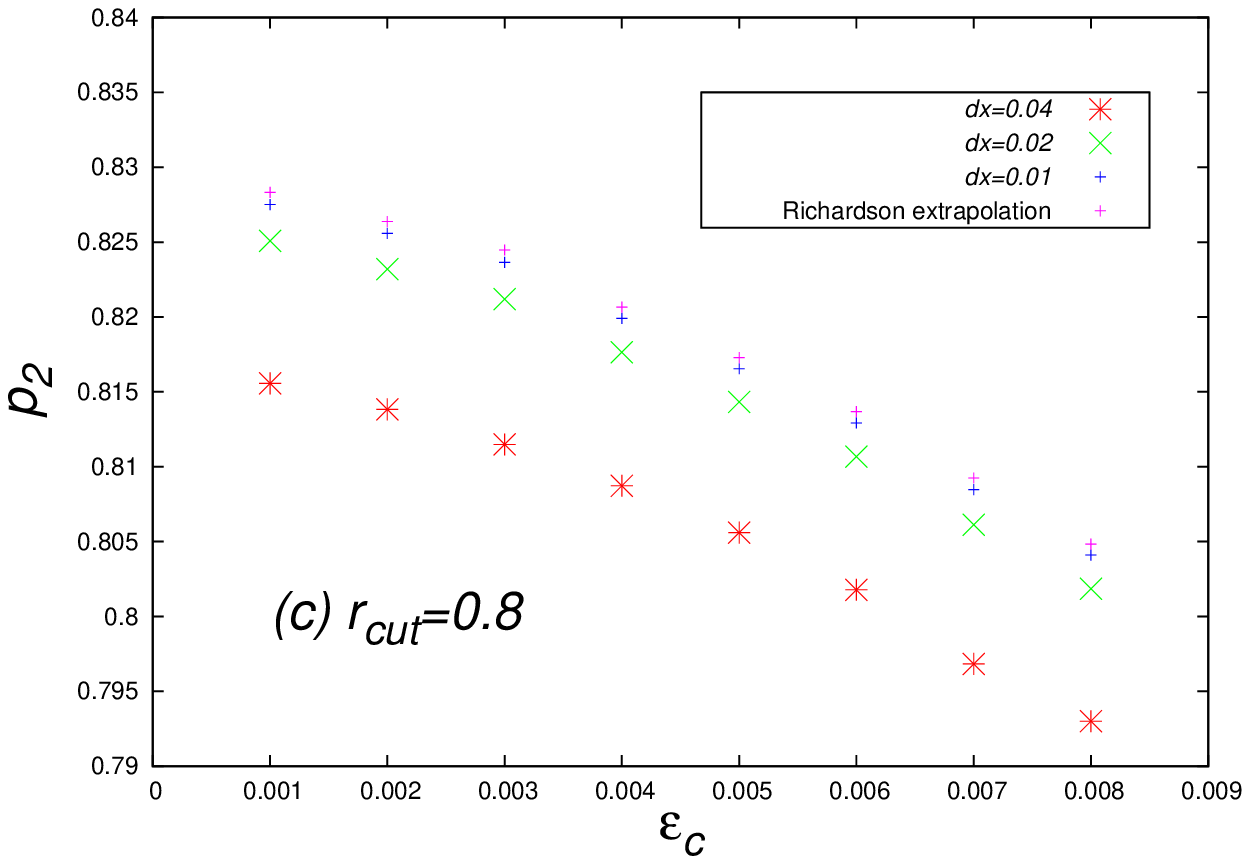}
\includegraphics[width=8cm]{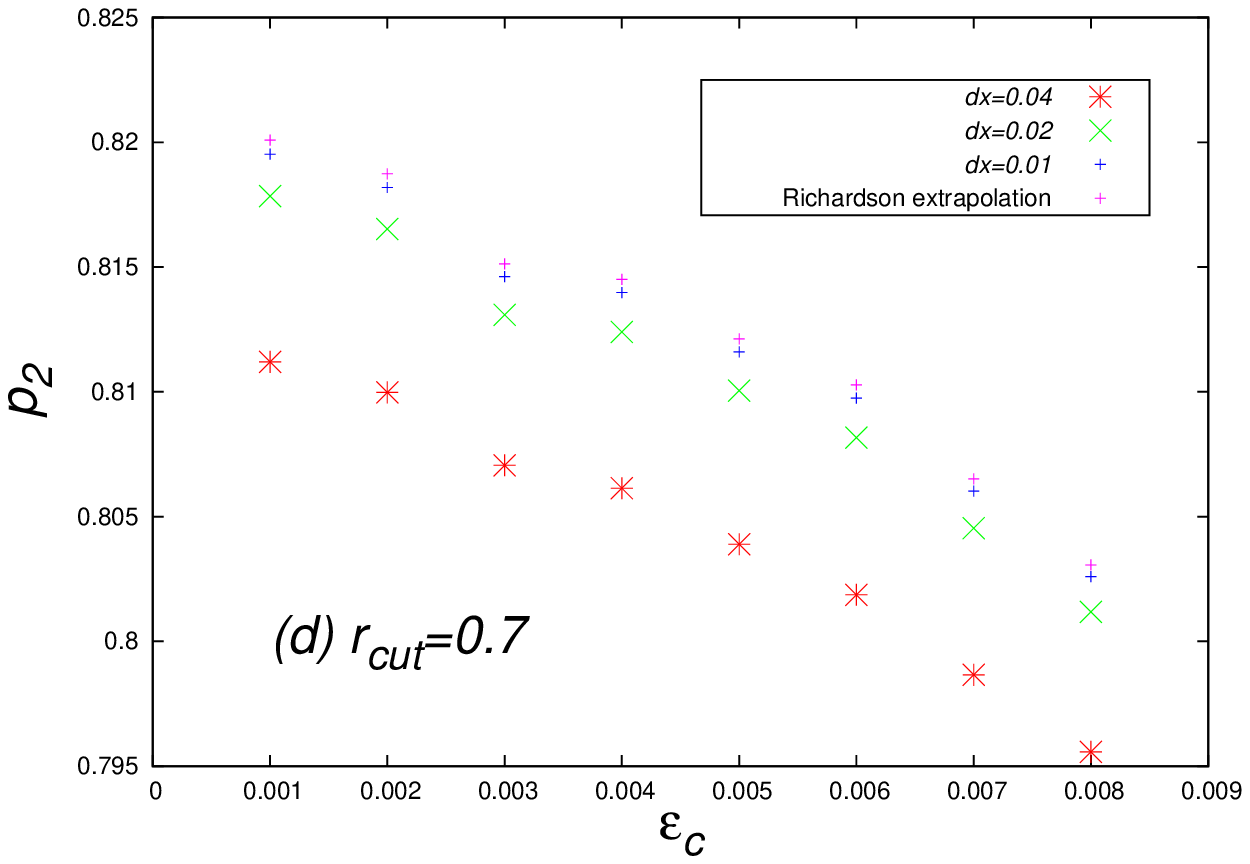}
\includegraphics[width=8cm]{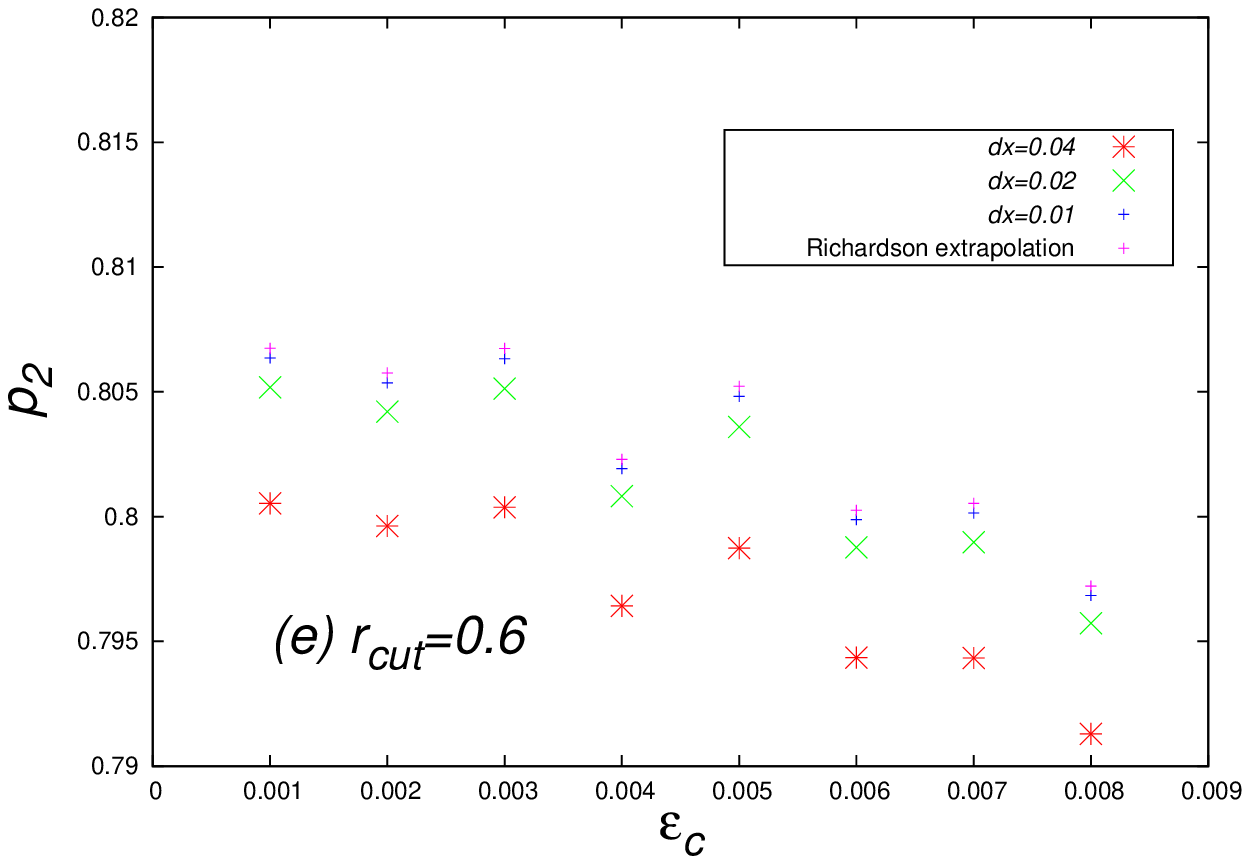}
\includegraphics[width=8cm]{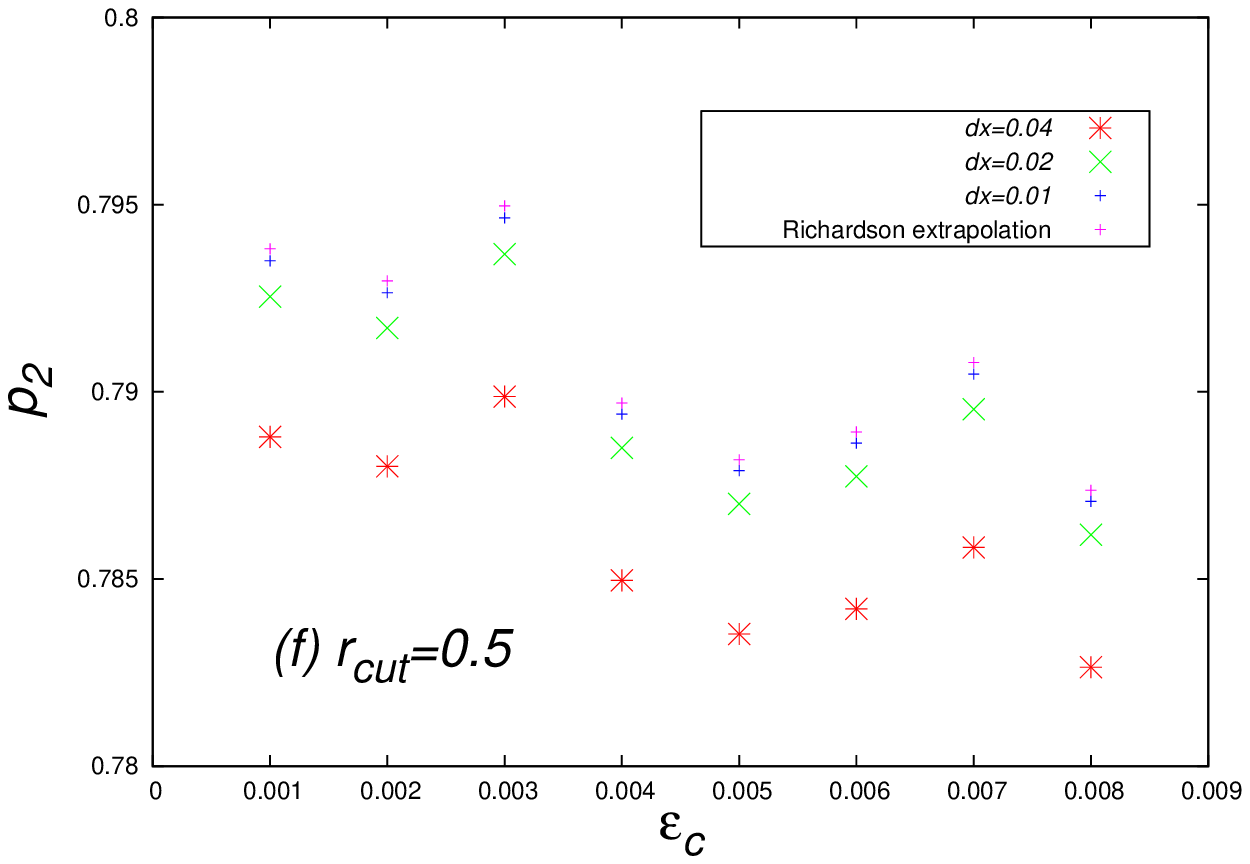}
\caption{\label{Fig:life-time_Symmetric} (a) We show a typical
evolution of the throat's areal radius $r_{throat}$ in terms of proper
time $\tau$ for the case of a collapse induced by a centered
perturbation. The inset shows the same function with a logarithmic
scale for $r_{throat}$ and illustrates how $r_{throat}$ converges to
zero in finite proper time $\tau$. Notice also that $\tau$ freezes
with respect to coordinate time which indicates that the lapse
collapses to zero near $x=0$. (b) The time scale $p_2$ is calculated
for different values of $\varepsilon_c$ and various resolutions using
$r_{cut}=0.9$. Second order convergence of the time scale for each
value of the amplitude is manifest. We predict the value at infinite
resolution using Richardson extrapolation. For this particular case,
the time scale for the smallest perturbation ($\varepsilon_c = 0.001$)
shows the highest time scale of all cases, namely $0.836$. This value
is comparable to the one obtained using linear perturbation theory
which is $0.846$, see Table I in Ref. \cite{jGfGoS-inprep1}. (c) to
(f) The same analysis for different values of $r_{cut}$. These plots
indicate a decreasing monotonic behaviour of the time scale as
$r_{cut}$ decreases, indicating that nonlinear terms tend to
accelerate the collapse.}
\end{figure*}

In figure~\ref{Fig:life-time_Symmetric} we show the results for the
time scale and its dependency on the cut-off value $r_{cut}$, the
initial parameter $\varepsilon_{c}$ and the resolution used for the
numerical evolutions. Two main features are: 1) The time scale for
small values of $\varepsilon_c$ and $r_{cut} \lesssim 1$ approach the
one calculated from linear perturbation theory \cite{jGfGoS-inprep1},
2) for large values of $\varepsilon_c$ and small values of $r_{cut}$,
the time scale is always smaller than the result of perturbation
theory, indicating that nonlinear terms tend to accelerate the
collapse.

\subsection{Formation of an apparent horizon}

\begin{figure}[htp]
\includegraphics[width=8cm]{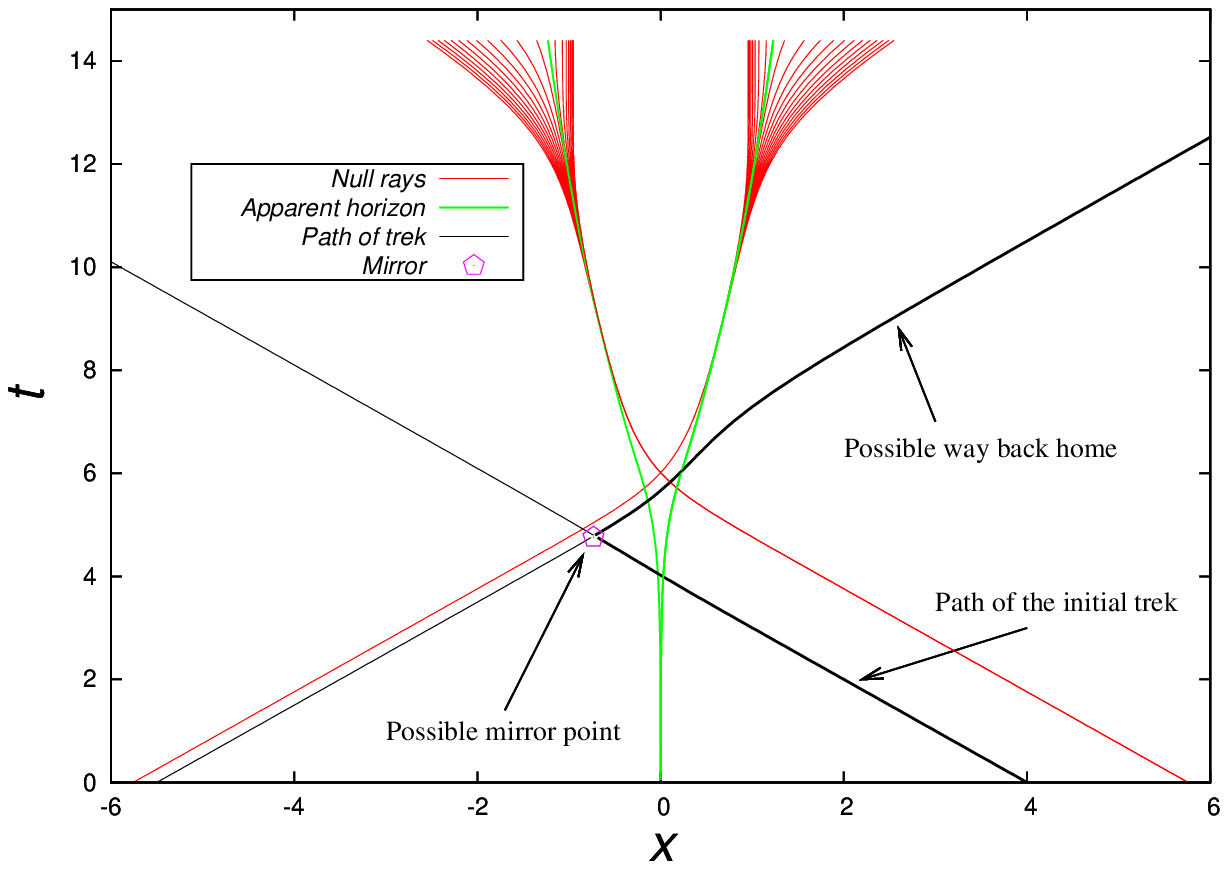}
\includegraphics[width=8cm]{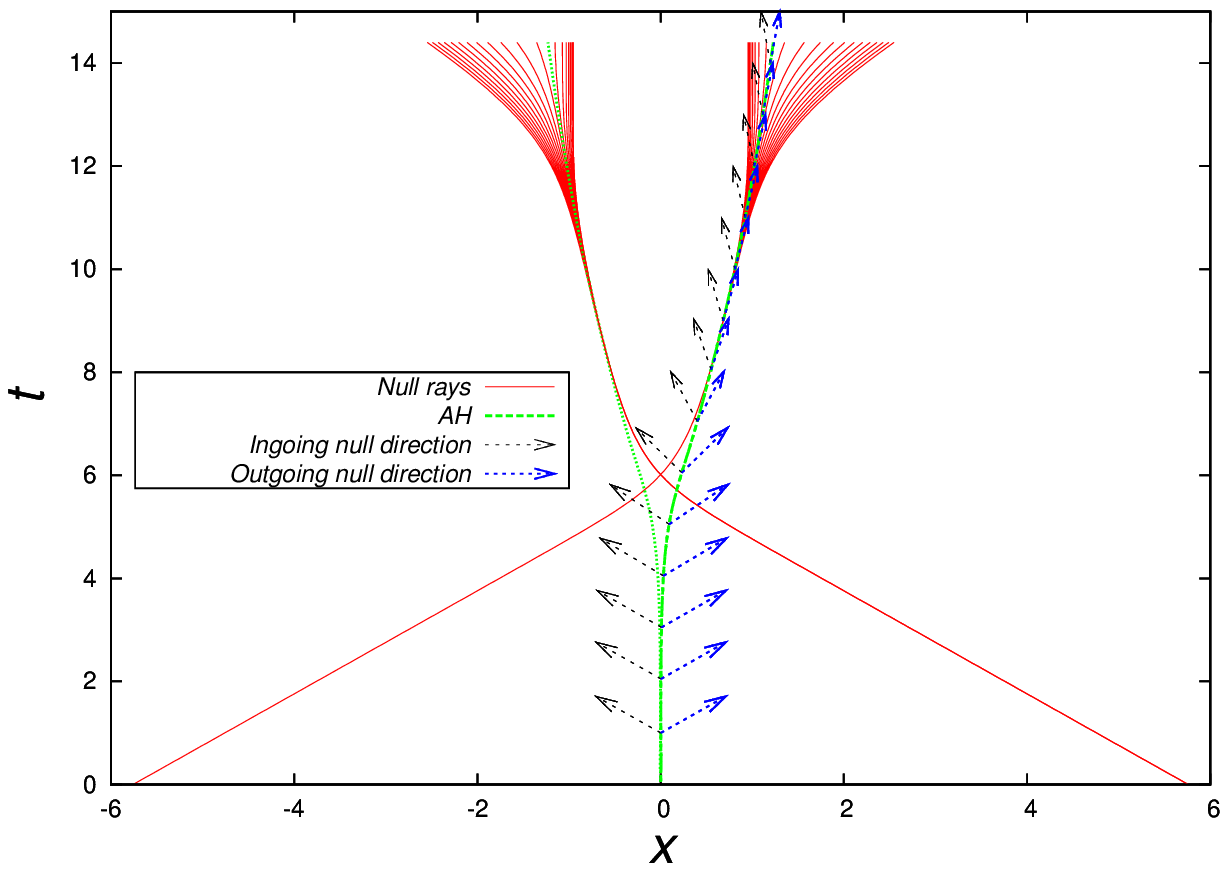}
\caption{\label{fig:geodesics_symmetric_pert} These plots correspond
  to the previously shown collapsing case. (Left panel) We show the
  apparent horizon surface and a bundle of null rays which depart from
  a surface identified with an approximation of an event horizon. The
  portion of the apparent horizon for $x > 0$ ($x < 0$) corresponds to
  the outermost marginally trapped surface as viewed from an observer
  at $x\to\infty$ ($x\to -\infty$). Correspondingly, the portion of
  the bundle of null rays emanating near $x = -6$ ($x = +6$)
  represents the event horizon with respect to the universe $x > 0$
  ($x < 0$). With respect to an observer in either universe, it can be
  seen that the apparent horizon lies {\em outside} the event
  horizon. The bold line indicates a possible trajectory of a null
  geodesic going from one universe to the other and reflected back.
  (Right panel) We show the future null directions at the location of
  the apparent horizon for $x>0$; the arrows departing from the
  apparent horizon indicate the in- and outgoing directions of light
  cones. This result indicates that the apparent horizon is actually a
  time-like surface.}
\end{figure}

During the collapse, we observe the formation of an apparent horizon.
This is shown in the left panel of
figure~\ref{fig:geodesics_symmetric_pert}, where we also show a bundle
of outgoing null rays whose intersection points with the initial slice
are fine-tuned in such a way that these rays lie as close as possible
to the apparent horizon for late times. As the plot suggests, this
bundle of light rays separates photons that might escape to future
null infinity from photons that are trapped inside a region of small
$x$. Therefore, we expect this bundle to represent a good
approximation for the event horizon of a black hole. As a consequence,
our simulations give a strong indication for the collapse of the
wormhole into a black hole.

The bold black line in the left panel of
figure~\ref{fig:geodesics_symmetric_pert} indicates a possible
trajectory of a photon traveling from one universe to the other and
back, allowing it to reach the asymptotic region of the home universe
without getting trapped by the black hole. In the right panel the
future null directions at the apparent horizon location are presented,
showing that the latter is a time-like surface.

\begin{figure}[htp]
\includegraphics[width=8cm]{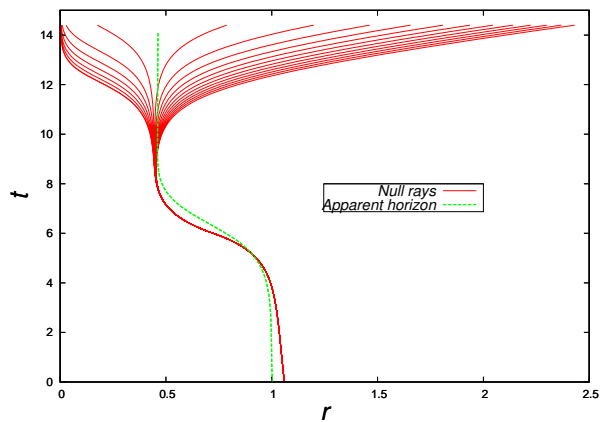}
\caption{\label{fig:geodesics_symmetric_pert_r} The trajectory of the
  apparent horizon and the bundle of outgoing null rays in the $t-r$
  diagram. Notice that the areal radius of both the apparent and the
  event horizons decreases in time. Even though these trajectories
  cross at $t \approx 5$ the event horizon lies {\em inside} the
  apparent horizon at all times, as is apparent from the $t-x$ diagram
  shown in figure~\ref{fig:geodesics_symmetric_pert} above.}
\end{figure}

Somehow unusual features of the collapse are that the apparent horizon
is not contained in the black hole region and that the areal radius of
both the apparent and event horizons {\em decreases} in time. This is
illustrated in figure~\ref{fig:geodesics_symmetric_pert_r}, where we
plot the trajectory of the apparent horizon and the bundle of outgoing
null rays in the $t-r$ diagram. Notice that for a spacetime satisfying
the null energy condition and cosmic censorship, propositions 9.2.1 in
\cite{HawkingEllis-Book} and 12.2.3 in \cite{Wald-Book} show that an
apparent horizon is contained in the black hole region. Under the same
assumptions theorem 12.2.6 in \cite{Wald-Book} and
\cite{sH71,pCeDgGrH01} establish that the area of the event horizon
cannot decrease in time.  In the present case, these results do not
apply since the null energy condition is not satisfied.  The decrease
of the area of the horizons could be related to the absorption of the
ghost scalar field by the black hole.

\subsection{The final state}

\begin{figure}[htp]
\includegraphics[width=8cm]{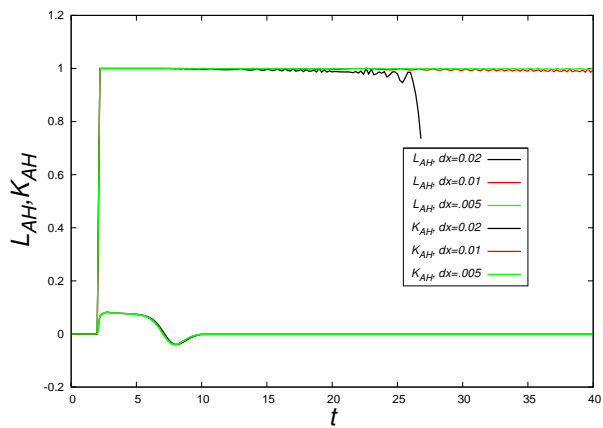}
\caption{\label{fig:invariants_collapsing} The scalars $L$ and $K$
evaluated at the apparent horizon versus coordinate time for different
resolutions. For high resolutions and late times, $L_{AH}$ converges
to zero and $K_{AH}$ to one. This indicates that the apparent horizon
converges to the event horizon of a Schwarzschild black hole. The
collapse is triggered by the off-centered perturbation with
$\varepsilon_c = 0.00125$, $\sigma_c=0.5$, $x_c = 2$.}
\end{figure}

In order to analyze the final state of the collapsing wormholes we
compute the scalars $L$ and $K=\frac{2M}{r}$ from
equations~(\ref{Eq:MInvariant},\ref{Eq:LInvariant}) at the apparent
horizon location. At the event horizon of a Schwarzschild black hole
these scalars are zero and one, respectively. As shown in
figure~\ref{fig:invariants_collapsing}, these quantities do indeed
converge to the corresponding Schwarzschild values for large times and
high spatial resolutions, indicating that the final state is a
Schwarzschild black hole, at least in the vicinity of the apparent
horizon.

\begin{figure}[ht]
\includegraphics[width=8cm]{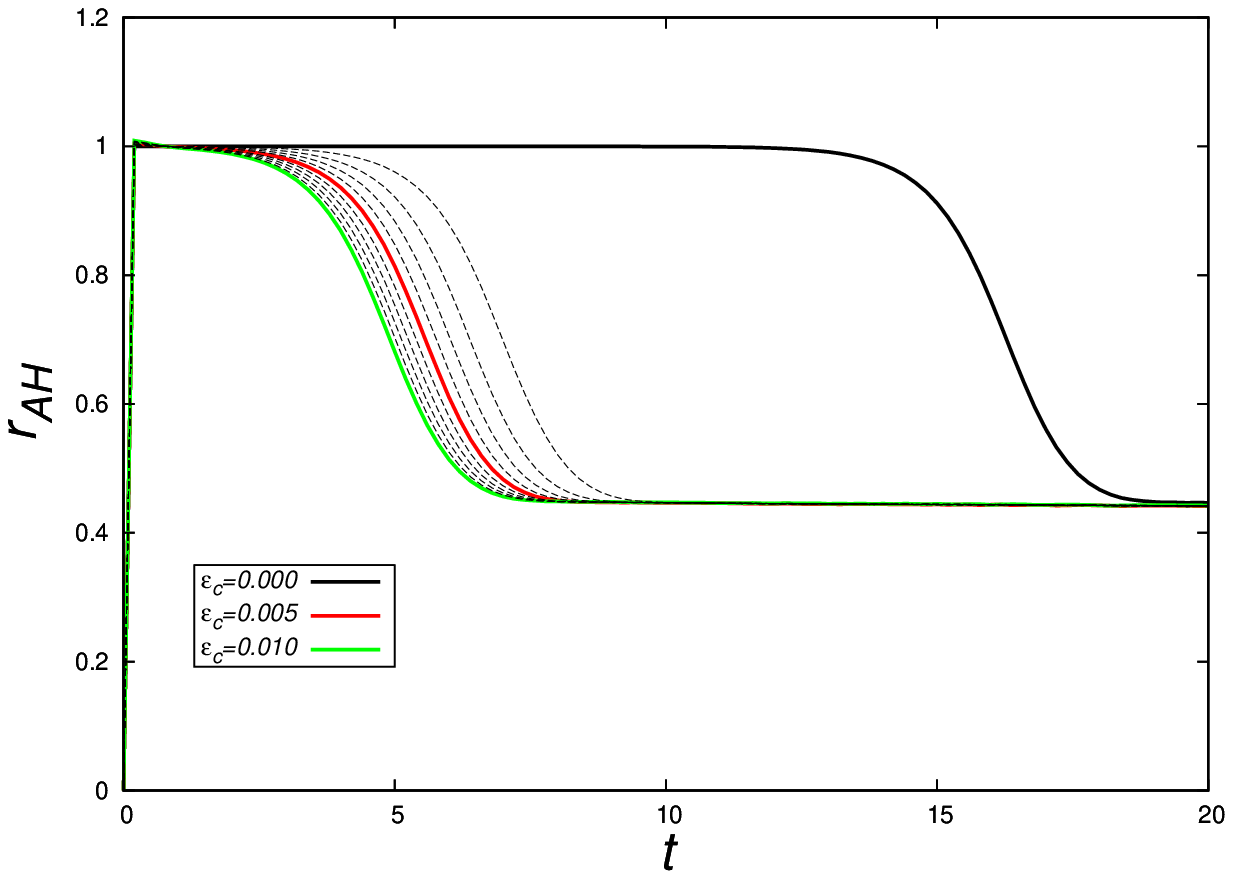}
\includegraphics[width=8cm]{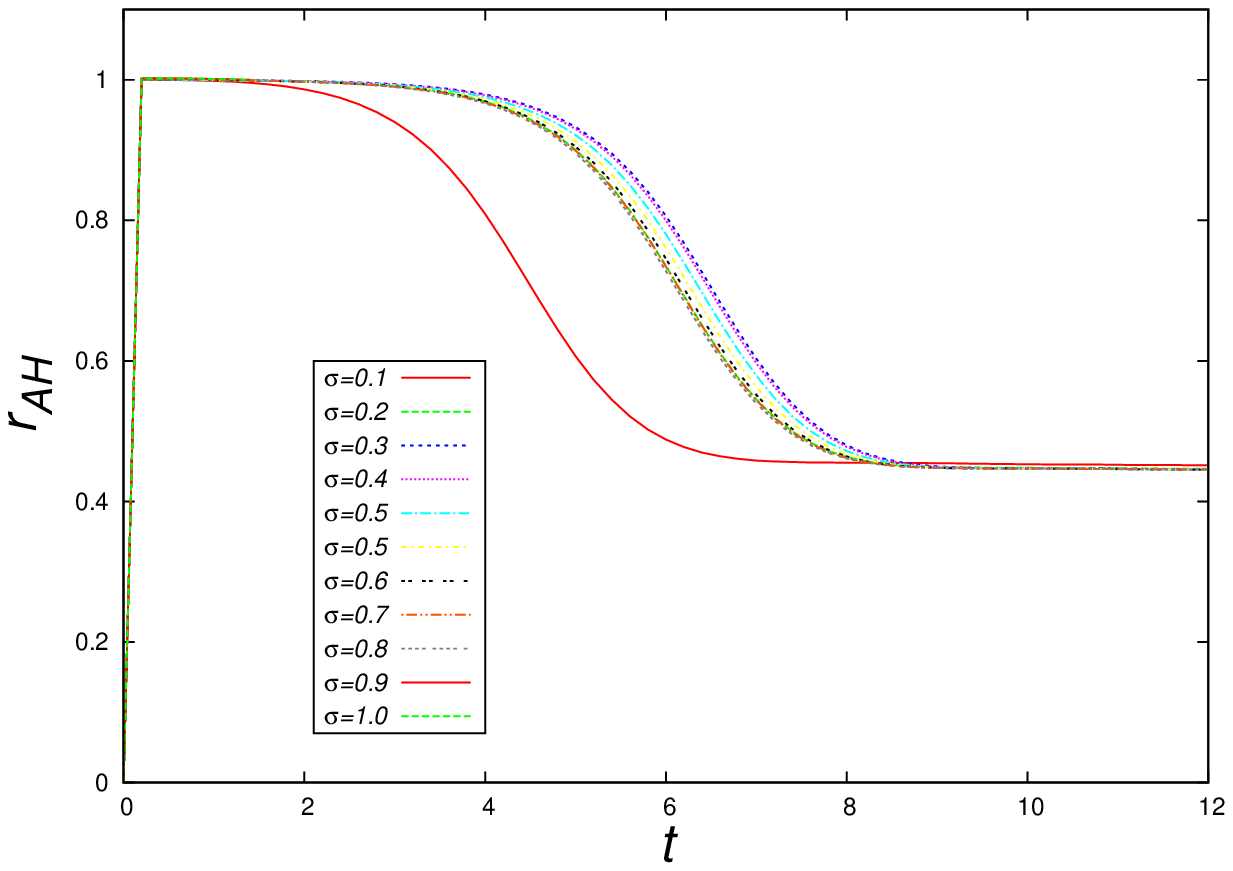}
\caption{Apparent horizon radius as function of the perturbations.
(Left) The width of the perturbation is fixed to $\sigma_c=0.5$ and
the amplitude of the perturbation ranges from $\varepsilon_c=0.000$ to
$\varepsilon_c=0.010$ in intervals of $\Delta\varepsilon_c=0.001$.
(Right) The amplitude of the perturbation is fixed to the value
$\varepsilon_c=0.002$ and the width varies from $\sigma_c=0.1$ to
$\sigma_c=1.0$ in intervals of $\Delta \sigma_c=0.1$. These plots
suggest that the final value for the areal radius of the apparent
horizon is independent of the perturbation in this range of parameter
space.}
\label{fig:r_ah_dependence}
\end{figure}

We try different initial parameters and perform runs with centered and
off-centered perturbations, and in all cases where the wormhole starts
collapsing the result is the same: an apparent horizon forms, and it
seems to settle down to the event horizon of a Schwarzschild black
hole. We also investigate the mass of the final black hole as a
function of the parameters of the perturbation. In
figure~\ref{fig:r_ah_dependence} we present the evolution of the
apparent horizon radius as a function of coordinate time for small,
positive values for the amplitude $\varepsilon_c$ and different values
for the width $\sigma_c$. For such values it seems that the final
black hole mass is universal and given by $m_{AH} \approx 0.22$. The
dependency of the black hole's final areal radius for a wider range of
initial parameters will be analyzed in section~\ref{sec:finalstate}.

\subsection{The distribution of the scalar field}

\begin{figure}[ht]
\includegraphics[width=8cm]{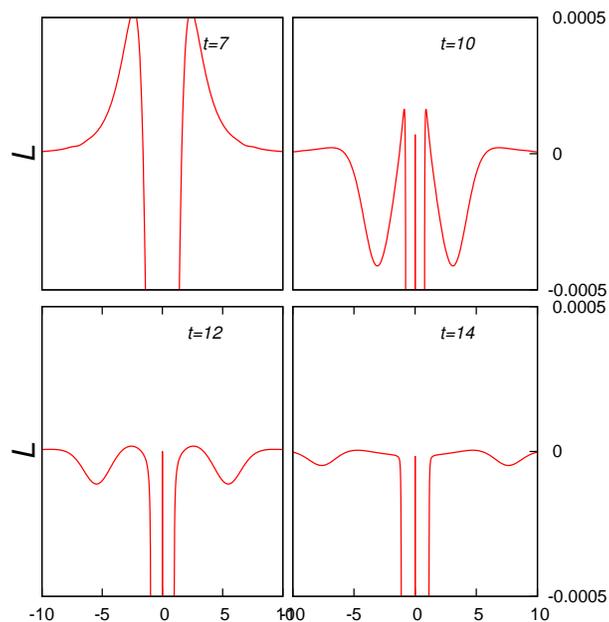}
\caption{Snapshots of $L$ for a zero mass collapsing
configuration and a centered perturbation. Notice the outgoing pulse 
with negative amplitude developing outside the apparent horizon region.}
\label{fig:L_snaps_collapse}
\end{figure}

In order to investigate the time evolution of the distribution of the
scalar field, we show in figure \ref{fig:L_snaps_collapse} snapshots
at different times of $L$ along the spatial domain for
the collapse of a zero mass configuration. A pulse of scalar field
departs from the location of the apparent horizon region and
propagates outwards with negative values. Since we know that the ADM
mass of the spacetime is zero and that the mass of the apparent
horizon is positive, we expect these pulses of scalar field to radiate
negative energy.

\begin{figure}[ht]
\includegraphics[width=8cm]{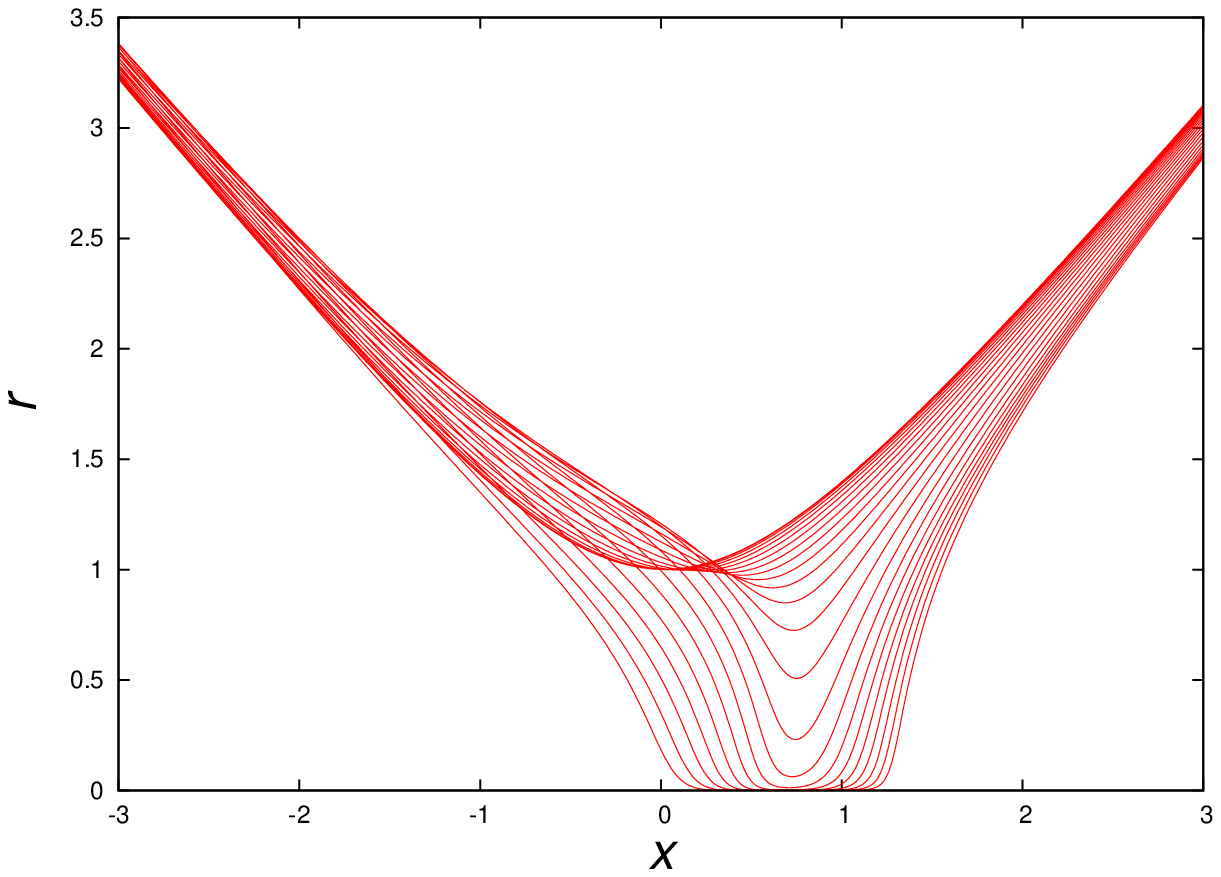}
\includegraphics[width=8cm]{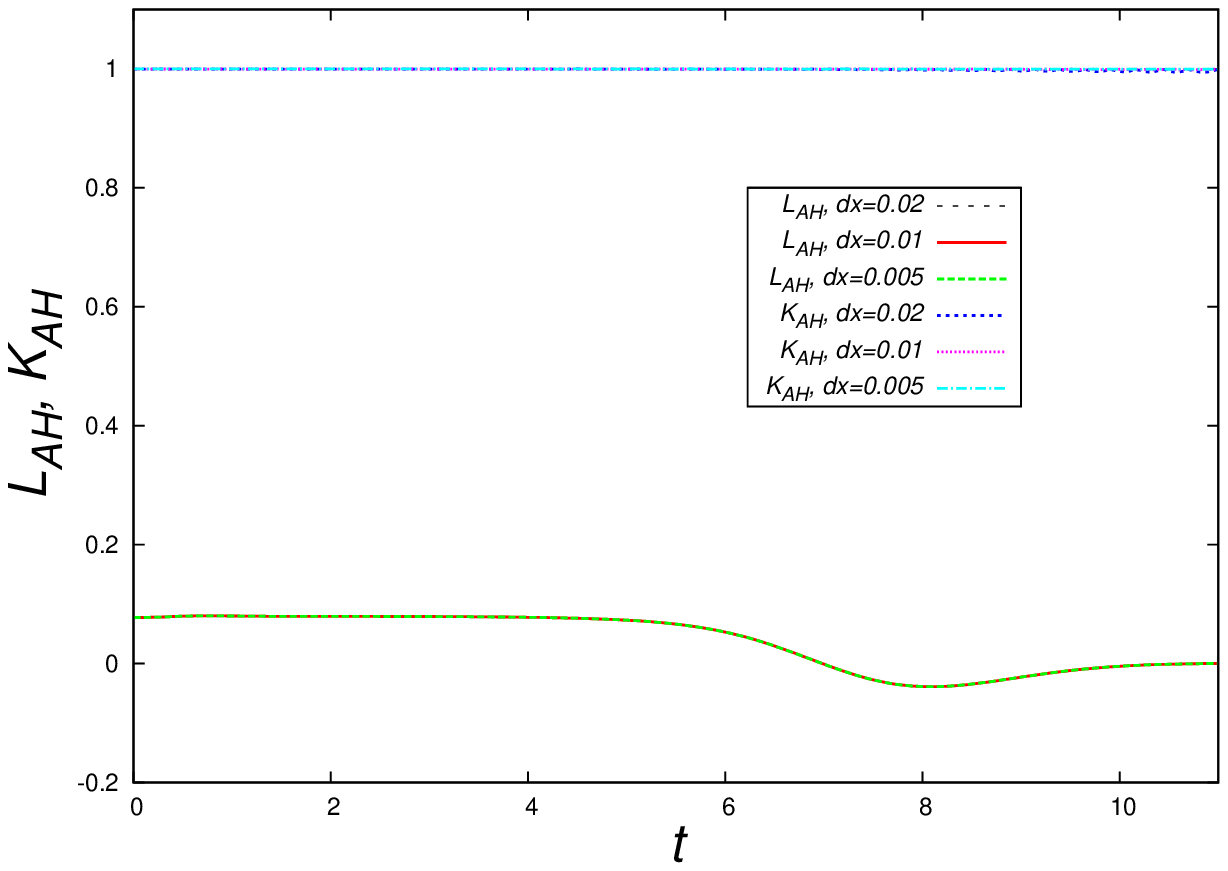}
\caption{(Left panel) Snapshots of the areal radius in terms of $x$
for several times. The particular behaviour of the massive case is
characterized by an initial boost of the solution in the coordinates
we use. This simulation was carried out using the physical parameter
$\gamma_1=0.015$ and the perturbation parameters
$\varepsilon_c=0.002$, $\sigma_c=0.5$ and $x_c=0$. (Right panel) The
values of $K$ and $L$ evaluated at the apparent
horizon. Once again we find that these scalars converge to their
Schwarzschild values for late times and high resolutions.}
\label{fig:massive_collpase_r}
\end{figure}

Finally, we also analyzed the collapse of massive wormholes which are
characterized by a positive value of the parameter $\gamma_1$. The
results are qualitatively similar to those found for the zero mass
case. For completeness, we show the behaviour of the areal radius for
a collapsing case and the scalars calculated at the apparent
horizon in figure~\ref{fig:massive_collpase_r}.

\section{The expanding case}
\label{sec:expanding}

\begin{figure*}[htp]
\includegraphics[width=8cm]{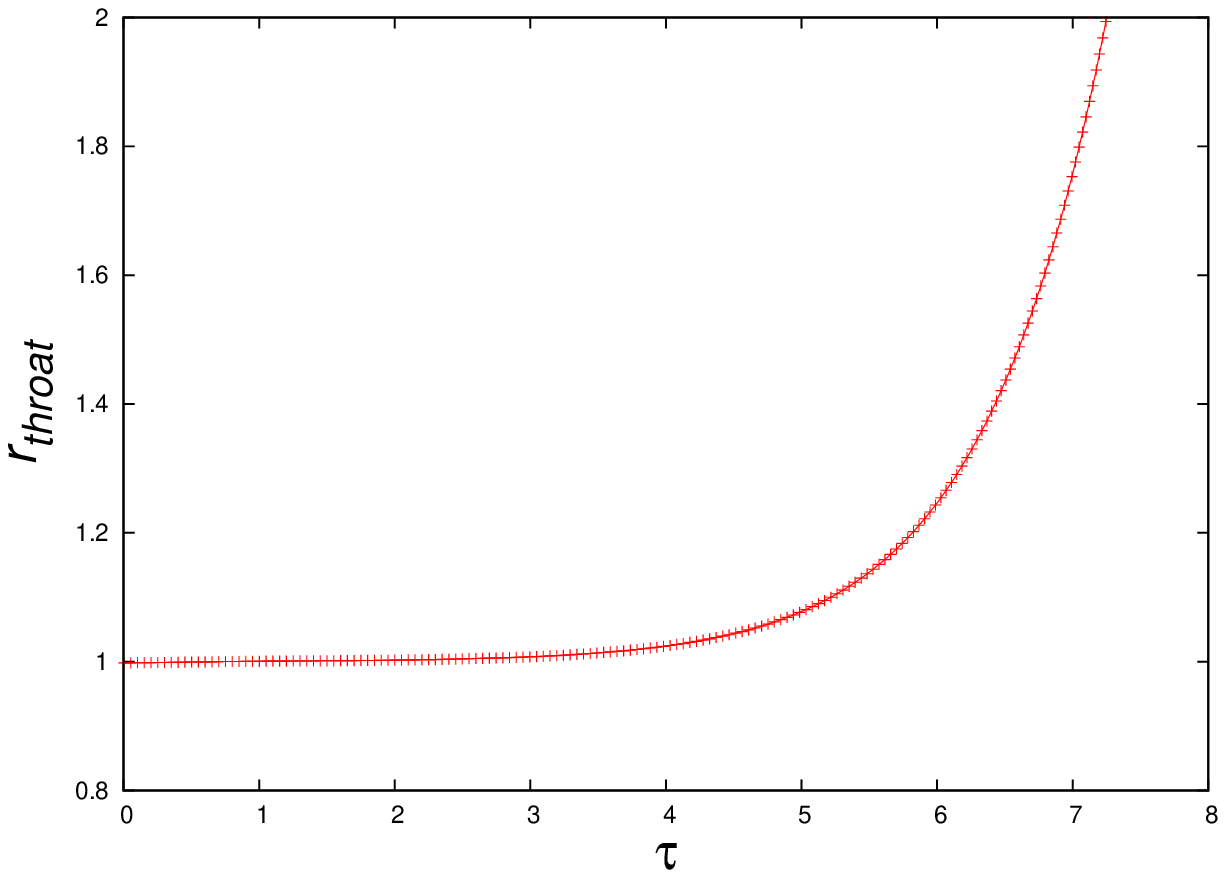}
\includegraphics[width=8cm]{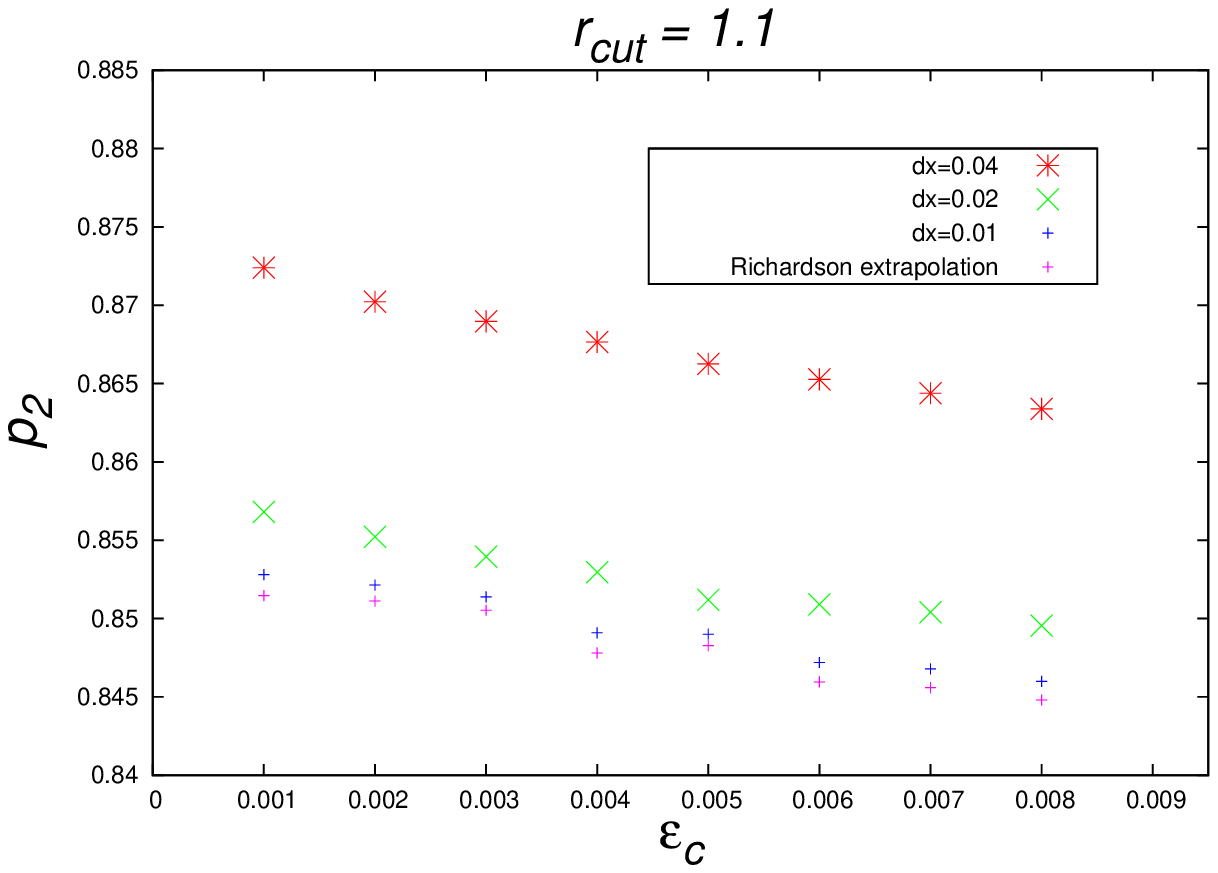}
\caption{\label{Fig:life-time_Symmetric_Expanding} (Left panel) We
show a typical evolution of the throat's areal radius versus proper
time $\tau$ for the expanding case under the action of a centered
perturbation. (Right panel) The time scale $p_2$ is calculated for
different values of $\varepsilon_c$ and various resolutions using
$r_{cut}=1.1$. Second order convergence of the time scale for each
value of the amplitude is manifest and the Richardson extrapolation
value is shown. The time scale for the smallest perturbation lies near
$0.85$.}
\end{figure*}

Perturbations that do not result in a collapse to a black hole induce
a rapid growth of the wormhole throat. In order to analyze this case,
we performed a series of simulations for the perturbed wormhole
solutions starting with the massless case $\gamma_1=0$. In
figure~\ref{Fig:life-time_Symmetric_Expanding} we consider the evolution
of a reflection symmetric perturbation and show the areal radius of
the spheres at the points of reflection symmetry $x=0$ as a function
of proper time. The exponential growth of this function indicates a
rapid growth of the wormhole.

\subsection{Time scale of the expansion}

Also shown in figure~\ref{Fig:life-time_Symmetric_Expanding} is the
time scale $p_2$ associated to the exponential growth for different
resolutions and different values of the initial amplitude
$\varepsilon_c$ of the perturbation. The function used to fit the
radius of the throat in this case is $r(\tau) = 1 +
e^{(\tau-p_1)/p_2}$. In the limit $\varepsilon_c \rightarrow 0$ the
time scale is estimated to be $0.85$ which is in good agreement with
the prediction from perturbation theory, see Table I in
Ref. \cite{jGfGoS-inprep1}.

\subsection{Slice-dependence of the throat}

\begin{figure}[ht]
\includegraphics[width=8cm]{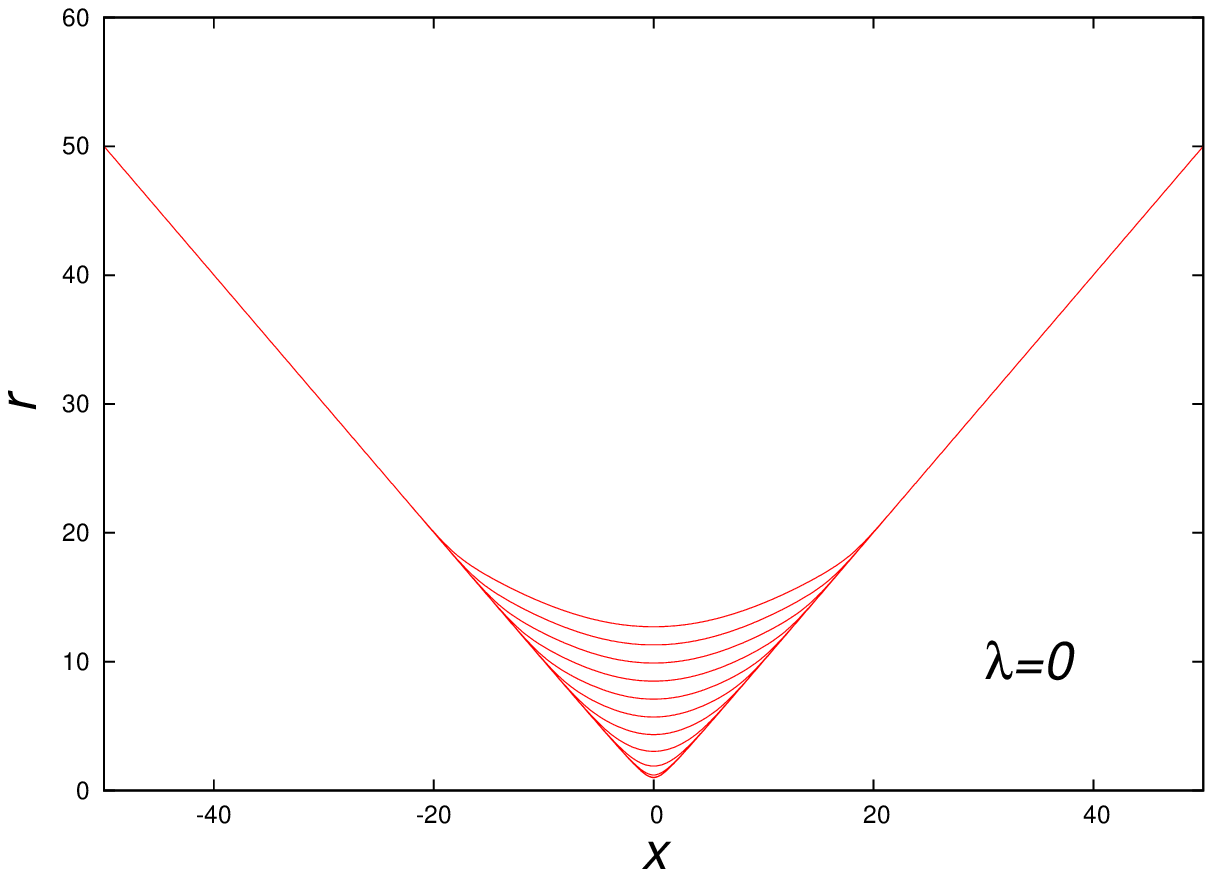}
\includegraphics[width=8cm]{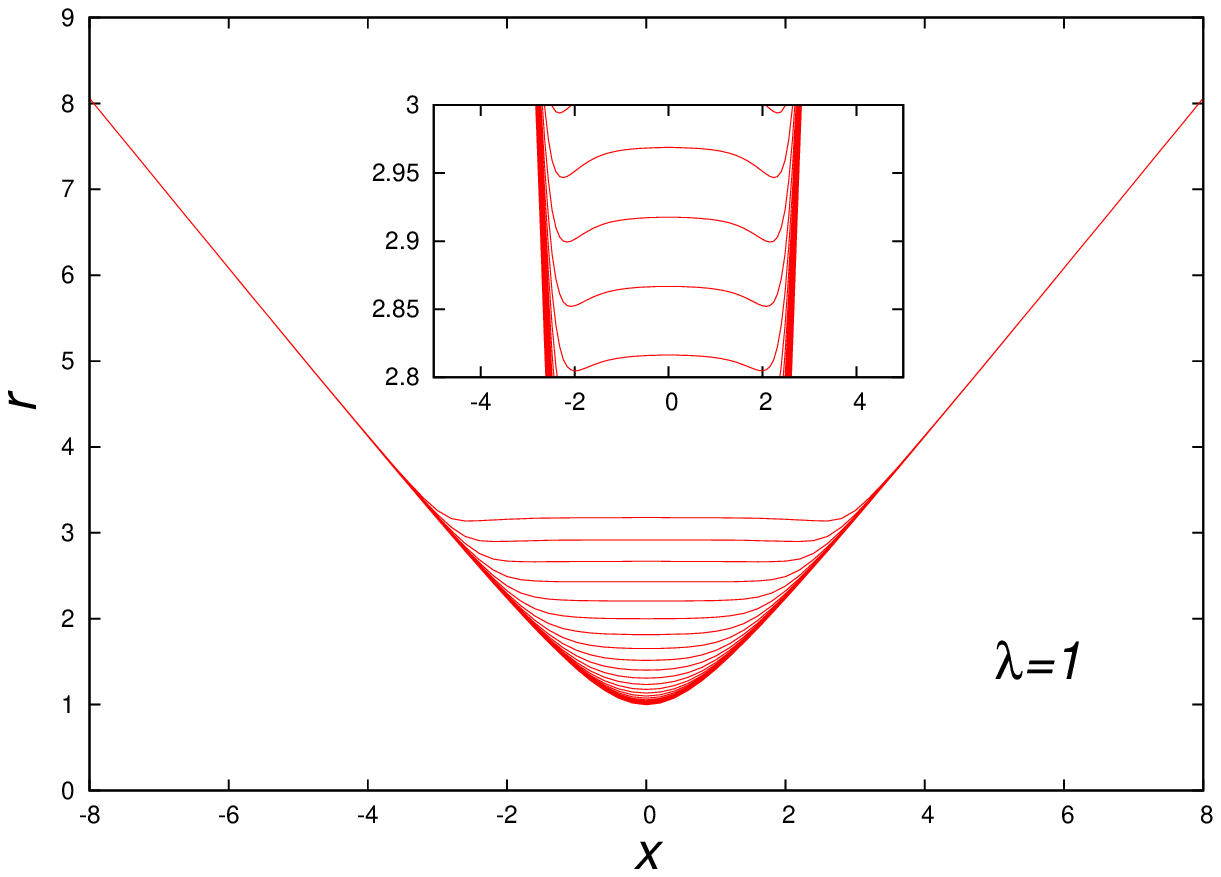}
\caption{Areal radius $r$ versus $x$ for $\lambda=0$ and $\lambda=1$
for the case of expansion using a centered perturbation. When
$\lambda=0$, the areal radius grows monotonically and preserves the
notion of a single throat in coordinate time. In the case $\lambda=1$
the areal radius develops a sort of two-throat image as can be seen in
the inset region of the plot.}
\label{fig:explosion}
\end{figure}

Next, we performed runs with the two gauge values $\lambda=0$ and
$\lambda=1$ with perturbations centered at the throat. While we did
not find the formation of any apparent horizons in any of these two
gauges, there is an interesting difference in the behaviour of the
areal radius $r$ as a function of the $x$ coordinate. This is shown in
figure~\ref{fig:explosion}. For $\lambda=0$ this function always has
its minimum at $x=0$, and one is tempted to define the throat's
location to be at $x=0$. However, looking at the results for
$\lambda=1$, one finds that although initially the minimum of $r$ as a
function of $x$ lies at $x=0$, at late times the function $r$
possesses a maximum at $x=0$ and two minima, one at a negative value
of $x$ and the other at a positive value of $x$. Therefore, from the
results with $\lambda=1$, one is led to the conclusion that {\em two}
throats develop.

\begin{figure}[ht]
\includegraphics[width=8cm]{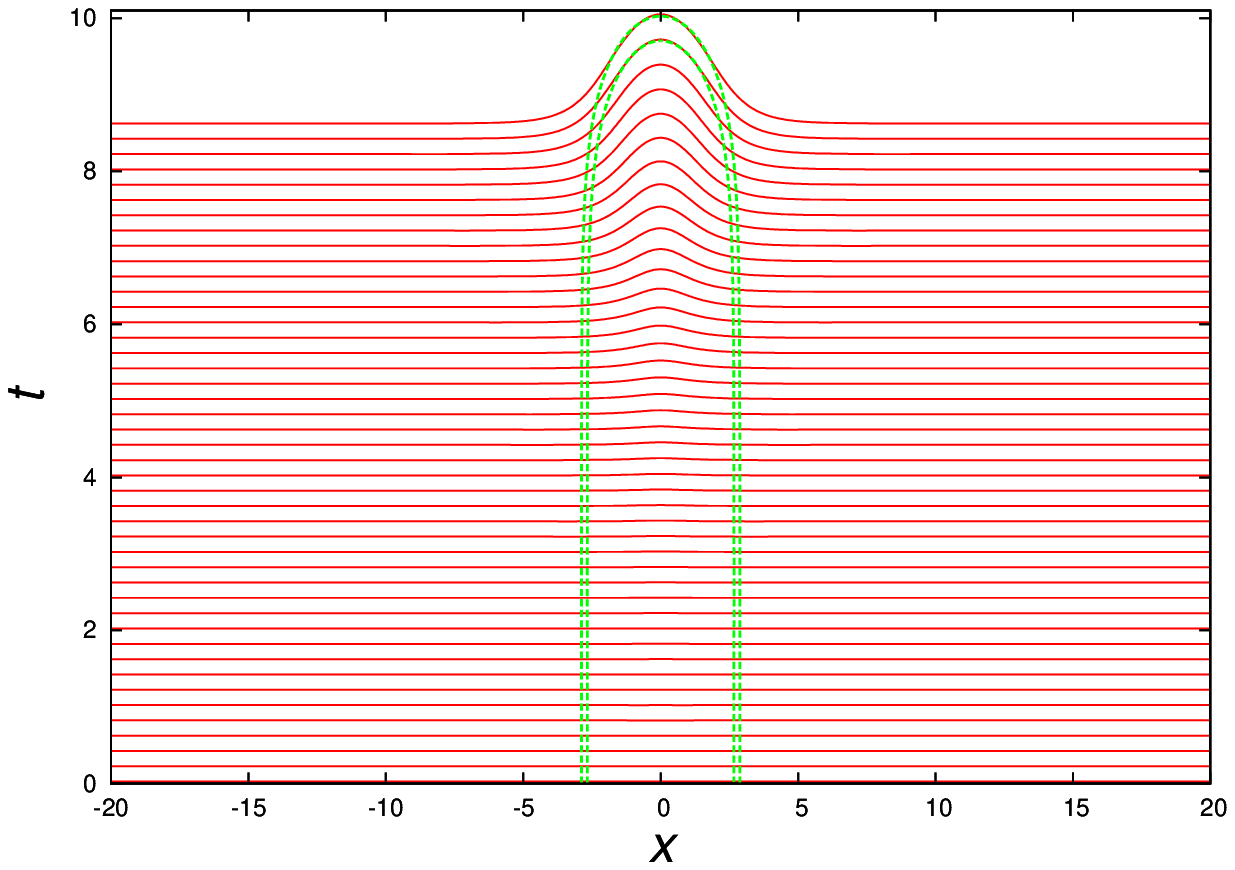}
\includegraphics[width=8cm]{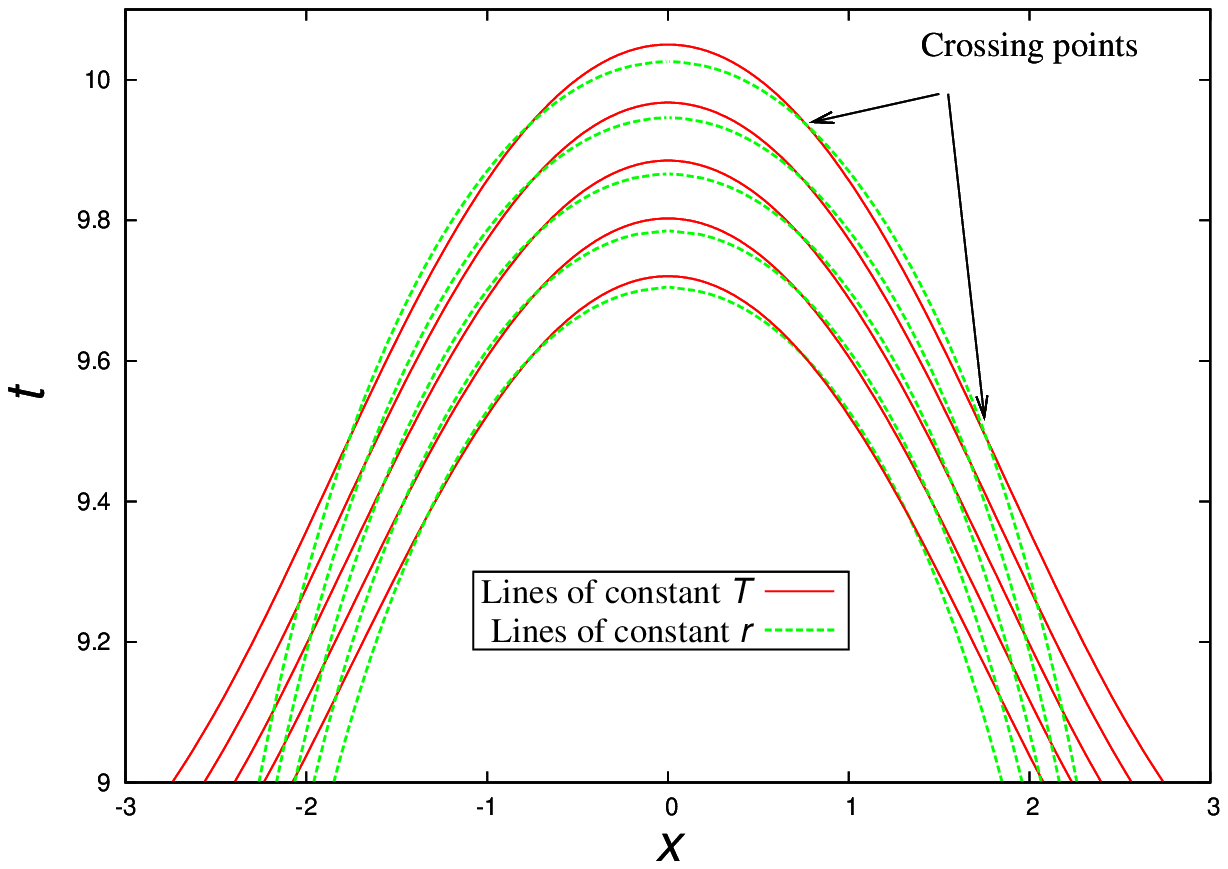}
\caption{Surfaces of constant $T$ (the time coordinate in the
coordinate system obtained from the evolution with $\lambda=1$) and
constant areal radius $r$. As can be seen, a $T=const$ surface may
cross a $r=const$ surface four times, indicating that the two-throat
image is a pure gauge effect.}
\label{fig:rconstante}
\end{figure}

It turns out that this apparent paradox is an effect of the different
slicing conditions resulting from the gauge condition
(\ref{Eq:SlicingCondition}) with $\lambda=0$ and $\lambda=1$,
respectively. In order to explain this, denote by $(t,x)$ the
conformal coordinates corresponding to the evolution with $\lambda=0$,
and let $(T,X)$ denote the coordinates obtained in the evolution with
$\lambda=1$. The relation between these two coordinate systems can be
obtained from the method described in section~\ref{subsec:ConfCoords}.
In figure~\ref{fig:rconstante} we show a spacetime diagram based on
the conformal coordinates $(t,x)$ where we plot the surfaces of
constant areal radius $r$ and the space-like slices of constant $T$.
As can be seen, at late enough times, the surfaces of constant $T$ may
cross the lines of constant areal radius four times. Therefore, as one
moves along a $T=const$ slice from one asymptotic end towards the
other, the areal radius decreases until a minimum is reached, then the
areal radius increases again until a local maximum is reached at
$x=0$; then $r$ decreases again until a local minimum is reached, and
finally the areal radius increases again. Therefore, the two-throat
image is a pure gauge effect, and the definition of the throat's
location as a local minimum of the areal radius over a given time
slice is not defined in a geometrical invariant way, but strongly
depends on the time foliation.

\subsection{The distribution of the scalar field}

\begin{figure}[ht]
\includegraphics[width=8cm]{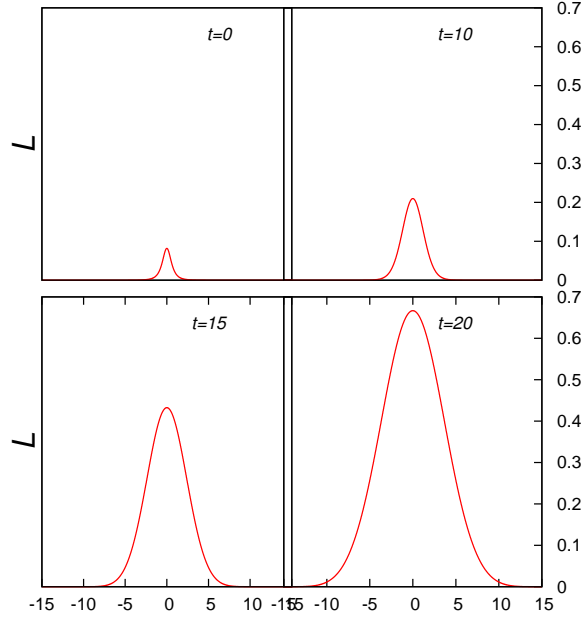}
\caption{Snapshots of $L$ for different times $t$ for
the case of an expanding wormhole with zero mass and centered 
perturbation.}
\label{fig:L_snaps_expand}
\end{figure}

In order to investigate the time evolution of the distribution of the
scalar field for the expanding case we show in
figure~\ref{fig:L_snaps_expand} snapshots at different times $t$ of
the scalar $L$ along the spatial domain for the expansion of a zero
mass wormhole. As can be seen from this plot, the amplitude of this
quantity increases in time while its shape remains essentially the
same. Therefore, in contrast to the collapsing case, the scalar field
does not seem to dissipate away but actually gains in strength near
the point of reflection symmetry $x=0$.

\begin{figure}[ht]
\includegraphics[width=8cm]{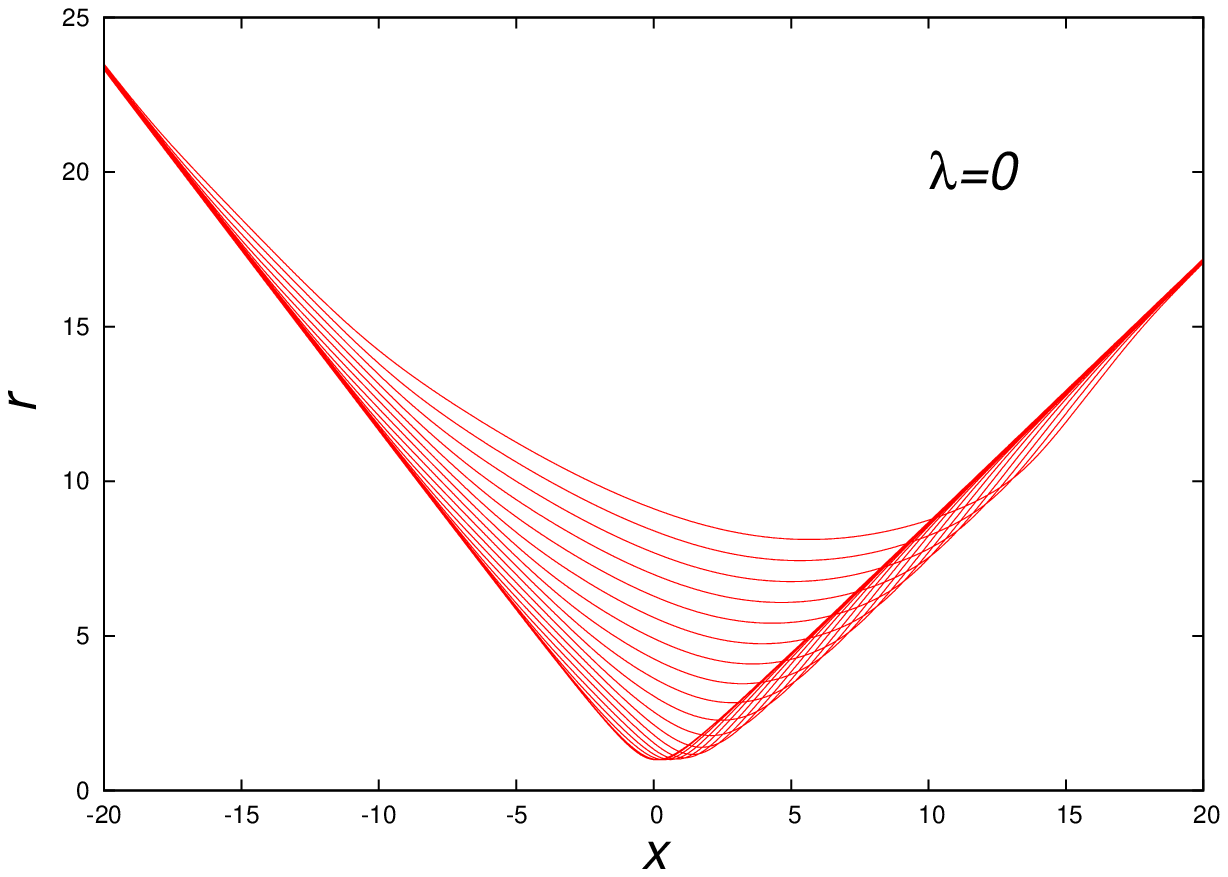}
\includegraphics[width=8cm]{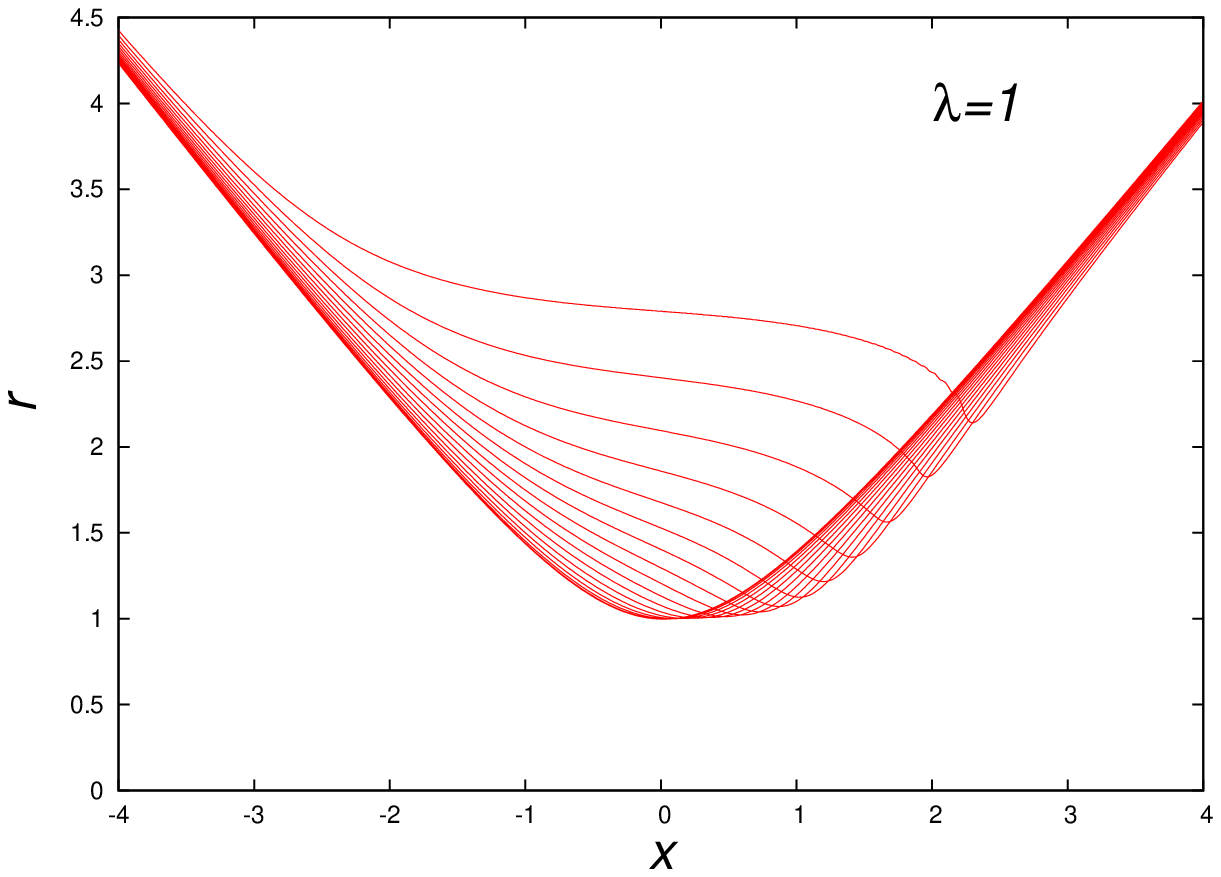}
\caption{Areal radius $r$ versus $x$ for $\lambda=0$ and $\lambda=1$
for a massive case. As in the collapsing case, the throat shifts
along the $x$ coordinate and then the areal radius expands. The
difference for different values of $\lambda$ is also manifest.}
\label{fig:expansion-massive}
\end{figure}

Finally, we also perform an evolution for a massive wormhole with 
$\gamma_1=0.015$. This is shown in figure~\ref{fig:expansion-massive}.

\section{Dependency of the final state on the initial perturbation}
\label{sec:finalstate}

\begin{figure}[ht]
\includegraphics[width=8cm]{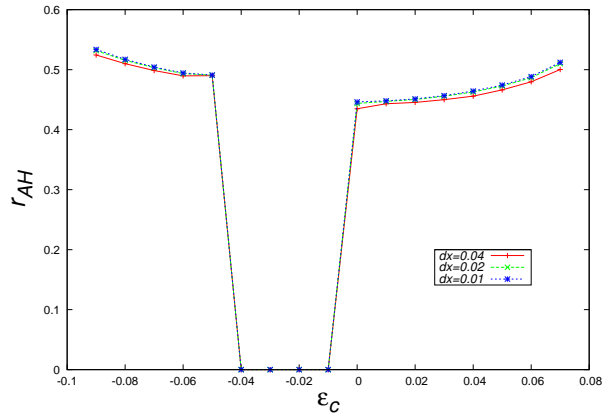}
\caption{The radius of the apparent horizon is shown for various
values of the amplitude of the perturbation. There are three regions
in the plot: i) one on the left, where a finite mass horizon forms,
ii) a second region in the middle for which there is expansion and
iii) another region for positive $\varepsilon_c$ for which there is
collapse again. For completeness, we also show these results for
various resolutions which indicate nearly second order convergence at
each point of the plot. The four points with $r_{AH}=0$ correspond to
expanding wormholes in which case there is no apparent horizon.}
\label{fig:r_ah_dependence2}
\end{figure}

We want to analyze the dependency of the final state of the system on
the initial perturbation. As we mentioned before, the two possible
final states of the evolution are the collapse into a black hole or
the expansion of the throat. In the former case, we can measure the
radius of the apparent horizon of the final black hole. In the latter,
there is no formation of black holes. We perform a systematic analysis
on the dependency of the apparent horizon's areal radius as a function
of the initial perturbation. As before, we perturb the function
$\bar{c}$ with a Gaussian pulse centered at $x_0=0$ with fixed width
$\sigma_c=0.5$, varying the amplitude $\varepsilon_c$ from
$\varepsilon_c \sim +0.1$ to $\varepsilon_c \sim -0.1$ in intervals of
$\Delta \varepsilon_c=0.01$.  When the values of the amplitude are
positive the system collapses into a black hole. We found that the
radius of the black hole decreases from $r_{AH} = 0.51$ for
$\varepsilon_c=0.07$ to $r_{AH} = 0.45$ for $\varepsilon_c=0$. Once
the values of the amplitude become negative the behaviour of the
evolution changes. In the interval $\varepsilon_c=[-0.04,0.00)$ the
wormhole expands instead of collapsing and no apparent horizon is
found.  It is tempting to stop here and assume that this behaviour
will continue.  Nevertheless, if we keep decreasing the value of the
amplitude we find that for values in the interval
$\varepsilon_c=[-0.09,-0.05]$ the system collapses again into a black
hole. In this range, the radius of the black hole increases from
$r_{AH}=0.49$ to $r_{AH}= 0.53$. These results are presented in
figure~\ref{fig:r_ah_dependence2} for three different resolutions in
order to emphasize the convergence of the apparent horizon's areal
radius.

\section{Conclusions}
\label{sec:conclusions}

Based on numerical methods, we analyzed in this work the nonlinear
stability of static, spherically symmetric general relativistic
wormhole solutions sourced by a massless ghost scalar field. This
complements the linear stability analysis performed in
\cite{jGfGoS-inprep1} where we proved that all such solutions are
unstable with respect to linear fluctuations. In particular, our
numerical simulations confirm the instability predicted by linear
theory and show that static, spherically symmetric wormholes sourced
by a massless ghost scalar field are also unstable with respect to
nonlinear fluctuations. Furthermore, we have checked that the time
scale associated to the instability agrees with the one computed from
perturbation theory, at least when computed over times for which the
departure from the equilibrium configuration is small.

Our numerical simulations also reveal that depending on the initial
perturbation, the wormholes either collapse to a Schwarzschild black
hole or undergo a rapid expansion. This confirms the results in
\cite{hSsH02} in the zero mass case and shows that a similar result
holds for massive wormholes. In the collapsing case, we reach this
conclusion by first observing the formation of an apparent horizon
whose areal radius converges to a fixed positive value at late
times. By computing geometrical quantities at the apparent
horizon and analyzing the behaviour of null geodesics in the vicinity
of the apparent horizon at late times we are led to the conclusion
that the final state of the collapse is a Schwarzschild black
hole. Further clues in support of this scenario are provided by the
observation that the scalar field disperses.

In the expanding case, the wormhole starts growing rapidly. For
reflection symmetric wormholes we find that the areal radius of the
spheres at the points of reflection symmetry grow exponentially with
respect to proper time. On the other hand, we have also found that the
definition of the throat as the three-surface obtained by determining
the global minimum of the areal radius in each time slice is
problematic in the sense that it may depend on the time foliation.

We have also performed a systematic study of the dependency of the
final state on the initial parameters of the perturbation. Using only
reflection symmetry initial data with a Gaussian perturbation profile
we investigated the possible universality of the final black hole's
mass. We found that for small, positive amplitudes of the initial
perturbation the final mass does not vary much with respect to the
initial width of the perturbation. For initial amplitudes which are
negative and small in magnitude, on the other hand, the wormholes
expand, the limiting case of zero amplitudes corresponding to a
threshold in parameter space separating collapsing wormholes from
expanding ones. By exploring a wide region of negative values for the
initial amplitude we also found a second threshold below which the
wormholes collapse again into a black hole.

Since a tiny perturbation may cause the wormhole to collapse into a
black hole, it is unlikely that the wormholes we have analyzed in this
article are useful for interstellar travel or building time
machines. Furthermore, the time scale associated to the collapse is of
the order of the throat's areal radius divided by the speed of light,
which is of the order of a few microseconds for a throat with areal
radius of one kilometer. We have analyzed the possible scenarios in
which a test particle may use the wormhole to tunnel from one universe
to the other and back before the black hole forms. As a consequence of
the instability, an observer moving on the path of such a test
particle cannot explore an arbitrarily large region of the other
universe if he wants to travel back home.


\acknowledgments

We thank Ulises Nucamendi and Thomas Zannias for many stimulating
discussions. This work was supported in part by grants CIC 4.9, 4.19
and 4.23 to Universidad Michoacana, PROMEP UMICH-PTC-121,
UMICH-PTC-195, UMICH-PTC-210 and UMICH-CA-22 from SEP Mexico and
CONACyT grant numbers 61173, 79601 and 79995.

\bibliographystyle{unsrt}
\bibliography{refs}
\end{document}